\documentstyle[a4,12pt]{article}
\begin{document}
\parskip=5pt plus 1pt minus 1pt

\begin{flushright}
{\large\bf DPNU-96-32}\\
{(hep-ph/9606422)}
\end{flushright}

\vspace{0.2cm}

\begin{center}
{\large\bf $D^0-\bar{D}^0$ Mixing and $CP$ Violation in Neutral $D$-meson Decays}
\end{center}

\vspace{0.3cm}
\begin{center}
{\sc Zhi-zhong Xing} \footnote{Electronic address:
xing@eken.phys.nagoya-u.ac.jp}
\end{center}

\begin{center}
{\it Department of Physics, Nagoya University, Chikusa-ku, Nagoya 464-01, Japan}
\end{center}

\vspace{2.5cm}

\begin{abstract}
$D^0-\bar{D}^0$ mixing at the detectable level or significant $CP$ violation in the 
charm system may strongly signify the existence of new physics. In view of the large
discovery potential associated with the fixed target experiments, the $B$-meson 
factories and the $\tau$-charm factories, we make a further study of the phenomenology
of $D^0-\bar{D}^0$ mixing and $CP$ violation in neutral $D$-meson decays. The generic
formulas for the time-dependent and time-integrated decay rates of both coherent and
incoherent $D^0\bar{D}^0$ events are derived, and their approximate expressions up to
the second order of the mixing parameters $x^{~}_D$ and $y^{~}_D$ are presented.
Explicitly we discuss $D^0-\bar{D}^0$ mixing and various $CP$-violating signals in
neutral $D$ decays to the semileptonic final states, the hadronic $CP$ eigenstates,
the hadronic non-$CP$ eigenstates and the $CP$-forbidden states. A few non-trivial 
approaches to the separate determination of $x^{~}_D$ and $y^{~}_D$ and to the 
demonstration of direct and indirect $CP$ asymmetries in the charm sector are suggested.
\end{abstract}

\vspace{1cm}

\begin{center}
[PACS number(s): 11.30.Er, 13.25.Ft, 14.40.Lb]
\end{center}

\newpage

\section{Introduction}

It is well known in particle physics that mixing between a neutral $P^0$ meson and
its $CP$-conjugate counterpart $\bar{P}^0$ can arise, if both of them couple to
a subset of virtual and (or) real intermediate states. Such mixing effects provide 
a mechanism whereby interference in the transition amplitudes of $P^0$ and $\bar{P}^0$
mesons may occur, leading to the possibility of $CP$ violation. Determining the
magnitude of $P^0-\bar{P}^0$ mixing and probing possible $CP$-violating phenomena
in the $P^0-\bar{P}^0$ system have been a challenging task for particle physicists.
To date, $K^0-\bar{K}^0$ and $B^0_d-\bar{B}^0_d$ mixing rates have been measured,
and the $CP$-violating signal induced by $K^0-\bar{K}^0$ mixing has been unambiguously
established \cite{PDG}. Many sophisticated experimental efforts, such as the programs of $\phi$
factories, $B$ factories and high-luminosity hadron machines, are being made to 
discover new signals of $CP$ asymmetries beyond the $K^0-\bar{K}^0$ system and to
precisely measure the Kobayashi-Maskawa (KM) matrix elements.

\vspace{0.3cm}

The study of mixing and $CP$ violation in the $Q=+2/3$ quark sector, particularly in the
$D^0-\bar{D}^0$ system, is not only complementary to our knowledge on the 
$K^0-\bar{K}^0$ and $B^0-\bar{B}^0$ systems, but also important for exploring
possible new physics that is out of reach of the standard model predictions. 
The rate of $D^0-\bar{D}^0$ mixing is commonly measured by two well-defined dimensionless 
parameters, $x^{~}_D$ and $y^{~}_D$, which correspond to the mass and width differences
of $D^0$ and $\bar{D}^0$ mass eigenstates. The latest E691 data of Fermilab fixed 
target experiments only give an upper bound on $D^0-\bar{D}^0$ mixing \cite{E691}:
\begin{equation}
r^{~}_D \; \approx \; \frac{x^2_D + y^2_D}{2} \; < \; 3.7 \times 10^{-3} \; .
\end{equation}
In the standard model the short-distance contribution to $D^0-\bar{D}^0$ mixing is via 
box diagrams and its magnitude is expected to be negligibly small 
($x^{\rm s.d.}_D \sim 10^{-5}$ and $y^{\rm s.d.}_D \leq x^{\rm s.d.}_D$ \cite{Cheng-Datta}) .
The long-distance effect on $D^0-\bar{D}^0$ mixing comes mainly from the
real intermediate states of $SU(3)$ multiplets, such as
\begin{equation}
D^0 \; \leftrightarrow \; \pi\pi \; , ~ \pi K \; , ~ \pi\bar{K} \; , ~
K\bar{K} \; \leftrightarrow \; \bar{D}^0 \; ,
\end{equation}
and is possible to be significant if the $SU(3)$ symmetry is badly
broken (e.g., $x^{\rm l.d}_D \sim y^{\rm l.d.}_D \sim 10^{-3} - 10^{-2}$ \cite{Wolfenstein-Kaeding}).
However, the dispersive approach \cite{Donoghue-Burdman} and the heavy quark effective 
theory \cite{Georgi} seem to favor a much smaller result for the long-distance contribution: 
$x^{\rm l.d.}_D \sim 10\times x^{\rm s.d.}_D$ and $x^{\rm l.d.}_D \sim x^{\rm s.d.}_D$
respectively. Such theoretical discrepancies indicate our poor understanding of the
dynamics for $D^0-\bar{D}^0$ mixing, hence more efforts in both theory and experiments
to better constrain the mixing rate are desirable. If calculations based on the standard
model can reliably limit $x^{~}_D$ and $y^{~}_D$ to be well below $10^{-2}$, then
observation of $r^{~}_D$ at the level of $10^{-4}$ or so will imply the existence of
new physics. On the other hand, improved experimental knowledge of $r^{~}_D$, in
particular the relative magnitude of $x^{~}_D$ and $y^{~}_D$, can definitely clarify 
the ambiguities in current theoretical estimates and shed some light on both the dynamics
of $D^0-\bar{D}^0$ mixing and possible sources of new physics beyond
the standard model.

\vspace{0.3cm}

The phenomenology of $CP$ violation in the $D^0-\bar{D}^0$ system was first developed 
by Bigi and Sanda \cite{Bigi-Sanda}, and further summarized by Bigi in ref. \cite{Bigi}. These works
have outlined the main features of $D^0-\bar{D}^0$ mixing and $CP$ asymmetries anticipated
to appear in neutral $D$-meson decays, although many of their formulas and results are
approximate or just for the illustrative purpose. 
The theoretical expectations on the magnitudes of various possible effects
are also sketched in refs. \cite{Bigi-Sanda,Bigi}.

\vspace{0.3cm}

Recent experimental progress, particularly in observing the doubly Cabibbo suppressed
decay (DCSD) $D^0\rightarrow K^+\pi^-$ \cite{CLEO}, constraining the $D^0-\bar{D}^0$ mixing rate \cite{PDG,E691}
and searching for the $CP$ asymmetries in $D$ decays to $K^+K^-$ etc \cite{E687-CLEO}, are quite
encouraging. Further experimental efforts, based mainly on the high-luminosity
fixed target facilities \cite{Kaplan}, the forthcoming $B$-meson factories and the proposed $\tau$-charm
factories \cite{Fry-Ruf}, are underway to approach the above physical goals. In view of the large discovery
potential associated with these experimental programs, a further study of the phenomenology
of $D^0-\bar{D}^0$ mixing and $CP$ violation in the charm system is no doubt necessary 
and important. 

\vspace{0.3cm}

In this paper we shall on the one hand follow the pioneering work of Bigi and Sanda to refine upon
the phenomenology of $D^0-\bar{D}^0$ mixing and $CP$ violation in neutral $D$ decays, and on
the other hand investigate some specific possibilities to separately determine $x^{~}_D$ and $y^{~}_D$
as well as to probe various $CP$-violating signals in the charm sector. A generic formulism
for the time-dependent and time-integrated decay rates of both coherent and incoherent
$D^0\bar{D}^0$ events are derived, and their approximate expressions up to $O(x^2_D)$ and
$O(y^2_D)$ are presented. Systematically but explicitly, we discuss a variety of 
$D^0-\bar{D}^0$ mixing and $CP$-violating measurables in neutral $D$ decays to the
semileptonic final states, the hadronic $CP$ eigenstates, the hadronic non-$CP$ eigenstates
and the $CP$-forbidden states. We show that it is possible to determine the relative magnitude
of $x^{~}_D$ and $y^{~}_D$ through observation of the dilepton events of coherent $D^0\bar{D}^0$
decays on the $\psi(4.14)$ resonance at a $\tau$-charm factory. A model-independent constraint on 
$D^0-\bar{D}^0$ mixing can also be obtained by measuring the decay-time distributions of 
$D^0/\bar{D}^0\rightarrow K_{S,L}+\pi^0$ etc. By use of the isospin analysis
and current data, we illustrate final-state interactions in $D\rightarrow K\bar{K}$ and
their influence on $CP$ violation. The interplay of $D^0-\bar{D}^0$ mixing and
DCSD effects in incoherent $D^0\bar{D}^0$ decays to $K^{\pm}\pi^{\mp}$ and in 
coherent $D^0\bar{D}^0$ decays to both $(l^{\pm}X^{\mp}, K^{\pm}\pi^{\mp})$
and $(K^{\pm}\pi^{\mp}, K^{\pm}\pi^{\mp})$ states is analyzed in the presence of $CP$ violation
and final-state interactions. We take a look at two types of $CP$-forbidden decays at the
$\psi(3.77)$ and $\psi(4.14)$ resonances. Finally the possibility to test the $\Delta Q =
\Delta C$ rule and $CPT$ symmetry in the $D^0-\bar{D}^0$ system is briefly discussed.

\vspace{0.3cm}

This work is organized as follows. In section 2 we derive the generic formulas for coherent
and incoherent $D^0\bar{D}^0$ decays, and then make some analytical approximations for them.
Sections 3, 4, 5 and 6 are devoted to $D^0-\bar{D}^0$ mixing and $CP$ violation in neutral
$D$ decays to the semileptonic states, the hadronic $CP$ eigenstates, the hadronic non-$CP$
eigenstates and the $CP$-forbidden states respectively, where some distinctive approaches
or examples are discussed for determining $x^{~}_D$ and $y^{~}_D$ or probing possible $CP$-violating
effects. We summarize our main results in section 7 with some comments on tests of the 
$\Delta Q = \Delta C$ rule and $CPT$ symmetry.

\section{Fundamental Formulas}
\setcounter{equation}{0}

We first develop a generic formulism for the time-dependent and time-integrated decays
of neutral $D$ mesons. Considering the smallness of $D^0-\bar{D}^0$ mixing indicated 
by both experimental searches and theoretical estimates, we then make some analytical
approximations for the obtained decay rates up to the accuracy of $O(x^2_D)$ and
$O(y^2_D)$.

\begin{center}
{\large\bf A. ~ Preliminaries}
\end{center}

In the assumption of $CPT$ invariance, the mass eigenstates of $D^0$ and $\bar{D}^0$ mesons
can be written as
\begin{eqnarray}
|D_{\rm L}\rangle & = & p |D^0\rangle ~ + ~ q |\bar{D}^0\rangle \; , \nonumber \\
|D_{\rm H}\rangle & = & p |D^0\rangle ~ - ~ q |\bar{D}^0\rangle \; ,
\end{eqnarray}
in which the subscripts ``L'' and ``H'' stand for Light and Heavy respectively,
and ($p, q$) are complex mixing parameters. Sometimes it is more convenient to
use the notation 
\begin{equation}
\frac{q}{p} \; \equiv \; \left | \frac{q}{p} \right | \exp({\rm i} 2\phi) \; ,
\end{equation}
where $\phi$ is a real $CP$-violating phase in $D^0-\bar{D}^0$ mixing. 
With the help of the conventions $CP|D^0\rangle = |\bar{D}^0\rangle$
and $CP|\bar{D}^0\rangle = |D^0\rangle$, the relations between 
the $CP$ eigenstates 
\begin{equation}
|D_1\rangle \; \equiv \; \frac{|D^0\rangle +|\bar{D}^0\rangle}{\sqrt{2}} \; , ~~~~~~~~~~
|D_2\rangle \; \equiv \; \frac{|D^0\rangle -|\bar{D}^0\rangle}{\sqrt{2}}
\end{equation}
and the mass eigenstates $|D_{\rm L}\rangle$, $|D_{\rm H}\rangle$ turn out to be
\begin{eqnarray}
|D_{\rm L}\rangle & = & \frac{p+q}{\sqrt{2}} |D_1\rangle ~ + ~ \frac{p-q}{\sqrt{2}} |D_2\rangle \; , \nonumber \\
|D_{\rm H}\rangle & = & \frac{p+q}{\sqrt{2}} |D_2\rangle ~ + ~ \frac{p-q}{\sqrt{2}} |D_1\rangle \; .
\end{eqnarray}
The proper-time evolution of an initially ($t=0$) pure 
$D^0$ or $\bar{D}^0$ is given as
\begin{eqnarray}
|D^0_{\rm phys}(t)\rangle & = & g^{~}_+(t) |D^0\rangle ~ 
+ ~ \frac{q}{p} g^{~}_-(t) |\bar{D}^0\rangle \; , \nonumber \\
|\bar{D}^0_{\rm phys}(t)\rangle & = & g^{~}_{+}(t)|\bar{D}^0\rangle ~ 
+ ~ \frac{p}{q} g^{~}_-(t) |D^0\rangle \; ,
\end{eqnarray}
where
\begin{eqnarray}
g^{~}_+(t) & = & \exp \left [ -\left ({\rm i} m +\frac{\Gamma}{2} \right ) t \right ] 
\cosh \left [ \left ( {\rm i} \Delta m - \frac{\Delta\Gamma}{2} \right ) \frac{t}{2} \right ] \; , \nonumber \\
g^{~}_-(t) & = & \exp \left [ -\left ({\rm i} m + \frac{\Gamma}{2} \right ) t \right ]
\sinh \left [ \left ( {\rm i} \Delta m - \frac{\Delta\Gamma}{2} \right ) \frac{t}{2} \right ] \; 
\end{eqnarray}
with the definitions
\begin{eqnarray}
m & \equiv & \frac{m^{~}_{\rm L}+m^{~}_{\rm H}}{2} \; , ~~~~~~~~~~
\Delta m \; \equiv \; m^{~}_{\rm H} - m^{~}_{\rm L} \; ; \nonumber \\
\Gamma & \equiv & \frac{\Gamma_{\rm L}+\Gamma_{\rm H}}{2} \; , ~~~~~~~~~~~~
\Delta\Gamma \; \equiv \; \Gamma_{\rm L} - \Gamma_{\rm H} \; .
\end{eqnarray}
Here $m^{~}_{\rm L(H)}$ and $\Gamma_{\rm L(H)}$ are the mass and width of $D_{\rm L(H)}$ respectively.
Note that the above definitions guarantee $\Delta m\geq 0$ and $\Delta\Gamma \geq 0$ in most cases. 
Practically, it is more popular to use the following two dimensionless parameters for 
describing $D^0-\bar{D}^0$ mixing:
\begin{equation}
x^{~}_D \; \equiv \; \frac{\Delta m}{\Gamma} \; , ~~~~~~~~~~ y^{~}_D \; \equiv \; \frac{\Delta\Gamma}{2\Gamma} \; .
\end{equation}
Certainly both $x^{~}_D$ and $y^{~}_D$ in most cases are positive (or vanishing).

\begin{center}
{\large\bf B. ~ Rates for incoherent $D$ decays}
\end{center}

The transition amplitude of a neutral $D$ meson decaying to a semileptonic or nonleptonic state $f$ can
be obtained from eq. (2.5) as follows:
\begin{eqnarray}
\langle f|{\cal H}|D^0_{\rm phys}(t)\rangle & = & g^{~}_+(t) A_f ~ 
+ ~ \frac{q}{p} g^{~}_-(t) \bar{A}_f \; , \nonumber \\
\langle f|{\cal H}|\bar{D}^0_{\rm phys}(t)\rangle & = & g^{~}_+(t) \bar{A}_f ~ 
+ ~ \frac{p}{q} g^{~}_-(t) A_f \; ,
\end{eqnarray}
where $A_f\equiv \langle f|{\cal H}|D^0\rangle$ and $\bar{A}_f \equiv \langle f|{\cal H}|\bar{D}^0\rangle$. 
For convenience, we also define the ratio of these two amplitudes:
\begin{equation}
\rho^{~}_f \; \equiv \; \frac{\bar{A}_f}{A_f} \; , ~~~~~~~~~~
\lambda_f \; \equiv \; \frac{q}{p}\rho^{~}_f \; .
\end{equation}
Then the time-dependent probabilities of such decay events are expressed as
\begin{eqnarray}
R(D^0_{\rm phys}(t)\rightarrow f) & \propto & |A_f|^2 \exp(-\Gamma t) \left [ C^{~}_y \cosh (y^{~}_D\Gamma t) 
~ + ~ C_x \cos (x^{~}_D\Gamma t) \right . \nonumber \\
& & ~~~~~~~~~~~~~~~~~ \left . ~ + ~ S^{~}_y \sinh (y^{~}_D\Gamma t) ~ 
+ ~ S_x \sin (x^{~}_D\Gamma t) \right ] \; , \nonumber  \\
R(\bar{D}^0_{\rm phys}(t)\to f) & \propto & |A_f|^2 \exp(-\Gamma t) \left [ \bar{C}^{~}_y \cosh (y^{~}_D\Gamma t) 
~ + ~ \bar{C}_x \cos (x^{~}_D\Gamma t) \right . \nonumber \\
& & ~~~~~~~~~~~~~~~~~ \left . ~ + ~ \bar{S}^{~}_y \sinh (y^{~}_D\Gamma t) ~ 
+ ~ \bar{S}_x \sin (x^{~}_D\Gamma t) \right ] \; , 
\end{eqnarray}
where 
\begin{eqnarray}
C_y & \equiv & \frac{1+|\lambda_f|^2}{2} \; , ~~~~~~~~~~
S_y \; \equiv \; -{\rm Re}\lambda_f \; , \nonumber \\
C_x & \equiv & \frac{1-|\lambda_f|^2}{2} \; , ~~~~~~~~~~
S_x \; \equiv \; -{\rm Im}\lambda_f \; ;
\end{eqnarray}
and
\begin{equation}
\left ( \bar{C}_y \; , ~ \bar{S}_y \; , ~ \bar{C}_x \; , ~ \bar{S}_x \right ) \; =\;
\left | \frac{p}{q} \right |^2 \left (C^{~}_y \; , ~ S_y \; , ~ - C^{~}_x \; , ~ - S_x \right ) \; .
\end{equation}
To obtain the time-independent decay rates, we integrate eq. (2.11) over $t\in [0, \infty)$ and
get
\begin{eqnarray}
R(D^0_{\rm phys}\rightarrow f) & \propto & |A_f|^2 \left [ \frac{1}{1-y^2_D}C_y + \frac{1}{1+x^2_D}C_x +
\frac{y^{~}_D}{1-y^2_D}S_y + \frac{x^{~}_D}{1+x^2_D}S_x \right ] \; , \nonumber \\
R(\bar{D}^0_{\rm phys}\rightarrow f) & \propto & |A_f|^2 \left [ \frac{1}{1-y^2_D}\bar{C}_y 
+ \frac{1}{1+x^2_D}\bar{C}_x +
\frac{y^{~}_D}{1-y^2_D}\bar{S}_y + \frac{x^{~}_D}{1+x^2_D}\bar{S}_x \right ] \; .
\end{eqnarray}
Eqs. (2.11) and (2.14) are the master formulas for incoherent $D$ decays.

\vspace{0.3cm}

Following the same way one can calculate the decay rates of $D^0_{\rm phys}$ and $\bar{D}^0_{\rm phys}$
to $\bar{f}$, the $CP$-conjugate state of $f$. To express the relevant formulas in analogy of eqs. (2.11)
and (2.14), we define
$\bar{A}_{\bar{f}}\equiv \langle \bar{f}|{\cal H}|\bar{D}^0\rangle$, $A_{\bar{f}}\equiv \langle 
\bar{f}|{\cal H}|D^0\rangle$ and 
\begin{equation}
\bar{\rho}^{~}_{\bar f} \; \equiv \; \frac{A_{\bar f}}{\bar{A}_{\bar f}} \; , ~~~~~~~~~~
\bar{\lambda}_{\bar{f}} \; \equiv \; \frac{p}{q}\bar{\rho}^{~}_{\bar f} \; .
\end{equation}
Then $R(D^0_{\rm phys}(t)\rightarrow \bar{f})$, $R(\bar{D}^0_{\rm phys}(t)\rightarrow \bar{f})$ 
and $R(D^0_{\rm phys}\rightarrow \bar{f})$, $R(\bar{D}^0_{\rm phys}\rightarrow \bar{f})$ can be
written out in terms of $\bar{A}_{\bar f}$ and $\bar{\lambda}_{\bar f}$. 
If $f$ is a $CP$ eigenstate (i.e., $|\bar{f}\rangle \equiv CP|f\rangle = \pm |f\rangle$), 
then we get $\bar{A}_{\bar{f}} = \pm \bar{A}_f$, $A_{\bar{f}} = \pm A_f$, $\bar{\rho}^{~}_{\bar f}
=1/\rho^{~}_f$ and $\bar{\lambda}_{\bar{f}}=1/\lambda_f$.

\begin{center}
{\large\bf C. ~ Rates for coherent $D$ decays}
\end{center}

For a coherent $D^0_{\rm phys}\bar{D}^0_{\rm phys}$ pair at rest, its time-dependent wave function
can be written as
\begin{equation}
\frac{1}{\sqrt{2}} \left [ D^0_{\rm phys}({\bf K},t)\rangle \otimes |\bar{D}^0_{\rm phys}(-{\bf K}, t)\rangle
~ + ~ C |D^0_{\rm phys}(-{\bf K}, t)\rangle \otimes |\bar{D}^0_{\rm phys}({\bf K}, t)\rangle \right ] \;  ,
\end{equation}
where $\bf K$ is the three-momentum vector of the $D$ mesons, and $C=\pm$ denotes the
charge-conjugation parity of this coherent system. The formulas for the time evolution of
$D^0_{\rm phys}$ and $\bar{D}^0_{\rm phys}$ mesons have been given in eq. (2.5). Here we consider
the case that one of the two $D$ mesons (with momentum $\bf K$) decays to a final state
$f_1$ at proper time $t_1$ and the other (with $-\bf K$) to $f_2$ at $t_2$. $f_1$ and
$f_2$ may be either hadronic or semileptonic states. The amplitude of such a
joint decay mode is given by
\begin{eqnarray}
A(f_1, t_1; f_2, t_2)_C & = & \frac{1}{\sqrt{2}} A_{f_1}A_{f_2} \xi_C
\left [ g^{~}_+(t_1)g^{~}_-(t_2) + C g^{~}_-(t_1)g^{~}_+(t_2) \right ] \; + \; \nonumber \\
& & \frac{1}{\sqrt{2}} A_{f_1}A_{f_2} \zeta_C 
\left [ g^{~}_+(t_1)g^{~}_+(t_2) + C g^{~}_-(t_1)g^{~}_-(t_2) \right ] \; ,
\end{eqnarray}
where $A_{f_i}\equiv \langle f_i|{\cal H}|D^0\rangle$ (with $i=1,2$), and
\begin{eqnarray}
\xi_C & \equiv & \frac{p}{q} \left ( 1 ~ + ~ C \lambda_{f_1} \lambda_{f_2} \right ) \; , \nonumber \\
\zeta_C & \equiv & \frac{p}{q} \left (\lambda_{f_2} ~ + ~ C \lambda_{f_1} \right ) \; .
\end{eqnarray}
Here the definition of $\lambda_{f_1}$ and $\lambda_{f_2}$ is similar to that of
$\lambda_f$ in eq. (2.10). After a lengthy calculation \cite{Xing1}, we obtain the time-dependent 
decay rate as follows:
\begin{eqnarray}
R(f_1, t_1; f_2, t_2)_C & \propto & |A_{f_1}|^2|A_{f_2}|^2 \exp (-\Gamma t_+) ~ \times  \nonumber \\
&  & \left [ \left (|\xi_C|^2 + |\zeta_C|^2 \right ) \cosh (y^{~}_D\Gamma t_C) 
-2 {\rm Re}\left (\xi^*_C \zeta_C \right ) \sinh (y^{~}_D\Gamma t_C) \right . \nonumber \\ 
&  & \left . - \left (|\xi_C|^2 - |\zeta_C|^2 \right ) \cos (x^{~}_D\Gamma t_C) 
+ 2{\rm Im}\left (\xi^*_C \zeta_C\right ) \sin (x^{~}_D\Gamma t_C) \right ] \; ,
\end{eqnarray}
where 
\begin{equation}
t_C \; \equiv \; t_2 ~ + ~ C t_1 
\end{equation}
has been defined.

\vspace{0.3cm}

The time-independent decay rate is obtainable from eq. (2.19) after the integration of $R(f_1, t_1; f_2, t_2)_C$
over $t_1\in [0, \infty)$ and $t_2\in [0, \infty)$:
\begin{eqnarray}
R(f_1, f_2)_C & \propto & |A_{f_1}|^2|A_{f_2}|^2 \left [ \frac{1+Cy^2_D}{(1-y^2_D)^2}
\left (|\xi_C|^2 + |\zeta_C|^2 \right ) ~ - ~ \frac{2(1+C)y^{~}_D}{(1-y^2_D)^2} {\rm Re}
\left (\xi^*_C \zeta_C \right ) \right . \nonumber \\ 
&  & \left . ~ - ~ \frac{1-Cx^2_D}{(1+x^2_D)^2} \left (|\xi_C|^2 - |\zeta_C|^2 \right ) 
~ + ~ \frac{2(1+C)x^{~}_D}{(1+x^2_D)^2} {\rm Im}\left (\xi^*_C \zeta_C\right ) \right ] \; .
\end{eqnarray}
We see that two interference terms ${\rm Re}(\xi^*_C\zeta_C)$ and ${\rm Im}(\xi^*_C\zeta_C)$
disappear in the case of $C=-1$, independent of the final states $f_1$ and $f_2$. 

\vspace{0.3cm}

In a similar way, one can calculate the joint decay rates of
$(D^0_{\rm phys}\bar{D}^0_{\rm phys})_C$ to $(f_1\bar{f}_2)$, $(\bar{f}_1f_2)$ or $(\bar{f}_1\bar{f}_2)$,
where $\bar{f}_1$ and $\bar{f}_2$ are $CP$-conjugate states of $f_1$ and $f_2$ respectively.

\begin{center}
{\large\bf D. ~ Analytical approximations}
\end{center}

In the standard model, the magnitudes of $x^{~}_D$ and $y^{~}_D$ are expected 
to be very small, at most of the order $10^{-2}$ 
(see, e.g., refs. \cite{Wolfenstein-Kaeding,Donoghue-Burdman,Georgi}). The current
experimental constraints on $D^0-\bar{D}^0$ mixing give $x^2_D+y^2_D<7.4\times 10^{-3}$
(see eq. (1.1)), which implies
$x^{~}_D<0.086$ and $y^{~}_D<0.086$. Due to the smallness of $x^{~}_D$ and $y^{~}_D$, 
the generic formulas obtained above
can be approximately simplified to a good degree of accuracy.

\vspace{0.3cm}

Up to the accuracy of $O(x^2_D)$ and $O(y^2_D)$ for every distinctive term, the time-dependent
decay rates in eq. (2.11) are approximated as 
\begin{eqnarray}
R(D^0_{\rm phys}(t)\rightarrow f) & \propto & |A_f|^2 \exp(-\Gamma t) \left [~ 1 ~ 
+ ~ \frac{1}{4} \left (x^2_D  +  y^2_D \right ) 
|\lambda_f|^2 \Gamma^2 t^2 \right . \nonumber \\
&  & \left .  - ~ \frac{1}{4} \left (x^2_D - y^2_D \right ) \Gamma^2 t^2 
~ - ~ \left (y^{~}_D {\rm Re}\lambda_f  ~ + ~ x^{~}_D {\rm Im}\lambda_f \right ) 
\Gamma t ~\right ] \; , \nonumber \\
R(\bar{D}^0_{\rm phys}(t)\rightarrow f) & \propto & |A_f|^2 \left | \frac{p}{q} \right |^2 
\exp(-\Gamma t) \left [ ~ |\lambda_f|^2 ~ + ~ \frac{1}{4} \left (x^2_D  +  y^2_D \right ) \Gamma^2 t^2  \right . \nonumber \\
&  & \left . - ~ \frac{1}{4} \left (x^2_D - y^2_D \right ) |\lambda_f|^2 \Gamma^2 t^2 
~ - ~ \left (y^{~}_D {\rm Re}\lambda_f  ~ - ~ x^{~}_D {\rm Im}\lambda_f \right ) \Gamma t ~\right ] \; .
\end{eqnarray}
Similarly we obtain the approximate decay rates for $D^0_{\rm phys}(t)\rightarrow \bar{f}$ and 
$\bar{D}^0_{\rm phys}(t)\rightarrow \bar{f}$:
\begin{eqnarray}
R(D^0_{\rm phys}(t)\rightarrow \bar{f}) & \propto & |\bar{A}_{\bar{f}}|^2 \left | \frac{q}{p} \right |^2
\exp(-\Gamma t) \left [~ |\bar{\lambda}_{\bar f}|^2  ~ + ~ \frac{1}{4} \left (x^2_D  +  y^2_D \right ) 
\Gamma^2 t^2  \right . \nonumber \\
&  & \left . - ~ \frac{1}{4} \left (x^2_D - y^2_D \right ) |\bar{\lambda}_{\bar f}|^2 \Gamma^2 t^2
~ - ~ \left (y^{~}_D {\rm Re}\bar{\lambda}_{\bar{f}}  ~ - ~ x^{~}_D {\rm Im}\bar{\lambda}_{\bar{f}} \right ) 
\Gamma t ~\right ] \; , \nonumber \\
R(\bar{D}^0_{\rm phys}(t)\rightarrow \bar{f}) & \propto & |\bar{A}_{\bar{f}}|^2 
\exp(-\Gamma t) \left [~ 1 ~ + ~ \frac{1}{4} \left (x^2_D  +  y^2_D \right ) 
|\bar{\lambda}_{\bar{f}}|^2 \Gamma^2 t^2  \right . \nonumber \\
&  & \left . - ~ \frac{1}{4} \left (x^2_D - y^2_D \right ) \Gamma^2 t^2 
~ - ~ \left (y^{~}_D {\rm Re}\bar{\lambda}_{\bar{f}}  ~ 
+ ~ x^{~}_D {\rm Im}\bar{\lambda}_{\bar{f}} \right ) \Gamma t ~\right ] \; .
\end{eqnarray}
The time-independent rates for these four processes turn out to be
\begin{eqnarray}
R(D^0_{\rm phys}\rightarrow f) & \propto & |A_f|^2 \left [~ 1 ~ 
+ ~ \frac{1}{2} \left (x^2_D + y^2_D \right ) |\lambda_f|^2 
\right . \nonumber \\
&  & \left . - ~ \frac{1}{2} \left (x^2_D - y^2_D \right ) ~ - ~ \left (y^{~}_D {\rm Re}\lambda_f  
+ x^{~}_D {\rm Im}\lambda_f \right ) 
~\right ] \; , \nonumber \\
R(\bar{D}^0_{\rm phys}\rightarrow \bar{f}) & \propto & |\bar{A}_{\bar f}|^2 
\left [~ 1 ~ + ~ \frac{1}{2} \left (x^2_D + y^2_D \right ) |\bar{\lambda}_{\bar f}|^2 \right . \nonumber \\
&  & \left . - ~ \frac{1}{2} \left (x^2_D - y^2_D \right ) ~ - ~ \left (y^{~}_D {\rm Re}\bar{\lambda}_{\bar f}  
+ x^{~}_D {\rm Im}\bar{\lambda}_{\bar f} \right ) ~\right ] \; ; 
\end{eqnarray}
and
\begin{eqnarray}
R(\bar{D}^0_{\rm phys}\rightarrow f) & \propto & |A_f|^2 \left | \frac{p}{q} \right |^2
\left [~ |\lambda_f|^2 ~ + ~ \frac{1}{2} \left (x^2_D + y^2_D \right ) \right . \nonumber \\
&  & \left . - ~ \frac{1}{2} \left (x^2_D - y^2_D \right ) |\lambda_f|^2 
~ - ~ \left (y^{~}_D {\rm Re}\lambda_f - x^{~}_D {\rm Im}\lambda_f \right ) ~\right ] \; , \nonumber \\
R(D^0_{\rm phys}\rightarrow \bar{f}) & \propto & |\bar{A}_{\bar{f}}|^2 \left | \frac{q}{p} \right |^2
\left [~ |\bar{\lambda}_{\bar f}|^2 ~ + ~ \frac{1}{2} \left (x^2_D + y^2_D \right ) \right . \nonumber \\
&  & \left . - ~ \frac{1}{2} \left (x^2_D - y^2_D \right ) |\bar{\lambda}_{\bar f}|^2 
~ - ~ \left (y^{~}_D {\rm Re}\bar{\lambda}_{\bar f} - x^{~}_D {\rm Im}\bar{\lambda}_{\bar f} \right ) ~\right ] \; .
\end{eqnarray}
The formulas listed above are very useful for the study of neutral $D$ decays in fixed target experiments or
at $B$-meson factories. Here no assumption has been made for the magnitudes of $|\lambda_f|$ and
$|\bar{\lambda}_{\bar f}|$. If they are considerably smaller than unity, e.g., in the DCSDs, then much
simpler expressions can be drawn from eqs. (2.22) to (2.25).

\vspace{0.3cm}

It is common knowledge that the decay-time distributions of coherent $(D^0_{\rm phys}\bar{D}^0_{\rm phys})_C$
pairs cannot be measured at a symmetric $e^+e^-$ collider \cite{Xing2}. Since the presently-proposed $\tau$-charm
factories are all based on symmetric $e^+e^-$ colisions, it is more practical to study the
time-integrated decays of $(D^0_{\rm phys}\bar{D}^0_{\rm phys})_C$ pairs. For completeness we shall present some
important formulas for the decay-time distributions of $(D^0_{\rm phys}\bar{D}^0_{\rm phys})_C$ events, with the
assumption of an asymmetric $\tau$-charm factory, in Appendix A. Such a work might be of purely 
academic sense, but it could also be useful in the future experiments of charm physics.

\vspace{0.3cm}

In the approximations up to $O(x^2_D)$ and $O(y^2_D)$, the time-integrated rates for
$(D^0_{\rm phys}\bar{D}^0_{\rm phys})_C$ decaying coherently
to $(f_1f_2)$, $(f_1\bar{f}_2)$, $(\bar{f}_1f_2)$ and
$(\bar{f}_1\bar{f}_2)$ states are obtained from eq. (2.21) as follows:
\begin{eqnarray}
R(f_1,f_2)_C & \propto & |A_{f_1}|^2 |A_{f_2}|^2 \left | \frac{p}{q} \right |^2 \left \{ (2+C) \left (x^2_D + y^2_D\right )
| 1+ C \lambda_{f_1}\lambda_{f_2}|^2 \right . \nonumber \\
&  & + ~ \left [ 2 - (2+C) \left (x^2_D - y^2_D \right ) \right ] |\lambda_{f_2} + C \lambda_{f_1}|^2 \nonumber \\
&  & - ~ 2(1+C) y^{~}_D \left [ \left (1+|\lambda_{f_1}|^2 \right ) {\rm Re}\lambda_{f_2} +
C \left (1+|\lambda_{f_2}|^2\right ) {\rm Re}\lambda_{f_1} \right ] \nonumber \\
&  & \left . + ~ 2(1+C) x^{~}_D \left [ \left (1-|\lambda_{f_1}|^2\right ) {\rm Im}\lambda_{f_2}
+ C \left (1-|\lambda_{f_2}|^2 \right ) {\rm Im}\lambda_{f_1} \right ] \right \} \; , \nonumber \\
R(\bar{f}_1,\bar{f}_2)_C & \propto & |\bar{A}_{\bar{f}_1}|^2 |\bar{A}_{\bar{f}_2}|^2 \left | \frac{q}{p} \right |^2 \left \{ (2+C) \left (x^2_D + y^2_D\right )
| 1+ C \bar{\lambda}_{\bar{f}_1}\bar{\lambda}_{\bar{f}_2}|^2 \right . \nonumber \\
&  & + ~ \left [ 2 - (2+C) \left (x^2_D - y^2_D \right ) \right ] |\bar{\lambda}_{\bar{f}_2} 
+ C \bar{\lambda}_{\bar{f}_1}|^2 \nonumber \\
&  & - ~ 2(1+C) y^{~}_D \left [ \left (1+|\bar{\lambda}_{\bar{f}_1}|^2 \right ) {\rm Re}\bar{\lambda}_{\bar{f}_2} +
C \left (1+|\bar{\lambda}_{\bar{f}_2}|^2\right ) {\rm Re}\bar{\lambda}_{\bar{f}_1} \right ] \nonumber \\
&  & \left . + ~ 2(1+C) x^{~}_D \left [ \left (1-|\bar{\lambda}_{\bar{f}_1}|^2\right ) {\rm Im}\bar{\lambda}_{\bar{f}_2}
+ C \left (1-|\bar{\lambda}_{\bar{f}_2}|^2 \right ) {\rm Im}\bar{\lambda}_{\bar{f}_1} \right ] \right \} \; ;
\end{eqnarray}
and
\begin{eqnarray}
R(f_1,\bar{f}_2)_C & \propto & |A_{f_1}|^2 |\bar{A}_{\bar{f}_2}|^2 \left \{ (2+C) \left (x^2_D + y^2_D\right )
| \bar{\lambda}_{\bar{f}_2} + C \lambda_{f_1}|^2 \right . \nonumber \\
&  & + ~ \left [ 2 - (2+C) \left (x^2_D - y^2_D \right ) \right ] | 1
+ C \lambda_{f_1} \bar{\lambda}_{\bar{f}_2}|^2 \nonumber \\
&  & - ~ 2(1+C) y^{~}_D \left [ \left (1+|\lambda_{f_1}|^2 \right ) {\rm Re}\bar{\lambda}_{\bar{f}_2} +
C \left (1+|\bar{\lambda}_{\bar{f}_2}|^2\right ) {\rm Re}\lambda_{f_1} \right ] \nonumber \\
&  & \left . - ~ 2(1+C) x^{~}_D \left [ \left (1-|\lambda_{f_1}|^2\right ) {\rm Im}\bar{\lambda}_{\bar{f}_2}
+ C \left (1-|\bar{\lambda}_{\bar{f}_2}|^2 \right ) {\rm Im}\lambda_{f_1} \right ] \right \} \; , \nonumber \\
R(\bar{f}_1,f_2)_C & \propto & |\bar{A}_{\bar{f}_1}|^2 |A_{f_2}|^2 \left \{ (2+C) \left (x^2_D + y^2_D\right )
|\lambda_{f_2} + C \bar{\lambda}_{\bar{f}_1}|^2 \right . \nonumber \\
&  & + ~ \left [ 2 - (2+C) \left (x^2_D - y^2_D \right ) \right ] |1
+ C \bar{\lambda}_{\bar{f}_1} \lambda_{f_2}|^2 \nonumber \\
&  & - ~ 2(1+C) y^{~}_D \left [ \left (1+|\bar{\lambda}_{\bar{f}_1}|^2 \right ) {\rm Re}\lambda_{f_2} +
C \left (1+|\lambda_{f_2}|^2\right ) {\rm Re}\bar{\lambda}_{\bar{f}_1} \right ] \nonumber \\
&  & \left . - ~ 2(1+C) x^{~}_D \left [ \left (1-|\bar{\lambda}_{\bar{f}_1}|^2\right ) {\rm Im}\lambda_{f_2}
+ C \left (1-|\lambda_{f_2}|^2 \right ) {\rm Im}\bar{\lambda}_{\bar{f}_1} \right ] \right \} \; .
\end{eqnarray}
Taking $f_1 = K^+l^-\bar{\nu}^{~}_l$ or $\bar{f}_1 = K^-l^+\nu^{~}_l$ for example, eqs. (2.26) and (2.27)
can be simplified significantly. Such semileptonic decay modes, which are flavor-specific, play the role in
identifying the flavor of the other $D$ meson decaying to $f_2$ or $\bar{f}_2$.

\section{Semileptonic $D$ decays}
\setcounter{equation}{0}

The manifestation of $D^0-\bar{D}^0$ mixing and $CP$ violation in the semileptonic decays of neutral $D$
mesons is relatively simple, since such transitions are flavor-specific in the standard model or some
of its extensions. Due to the flavor specification of $D^0\rightarrow l^+X^-$ and $\bar{D}^0\rightarrow
l^-X^+$, it is not necessary to study the time dependence of $D^0_{\rm phys}$ and $\bar{D}^0_{\rm phys}$
decay modes.

\begin{center}
{\large\bf A. ~ $D^0-\bar{D}^0$ mixing and $CP$ violation}
\end{center}

For fixed target experiments or $e^{+}e^{-}$ collisions at the $\Upsilon(4S)$ resonance,
the produced $D^0$ and $\bar{D}^0$ mesons are incoherent. Knowledge of $D^0-\bar{D}^0$
mixing is expected to come from ratios of the wrong-sign to right-sign events of
semileptonic $D$ decays:
\begin{equation}
r \; \equiv \; \frac{R(D^0_{\rm phys}\rightarrow l^-X^+)}{R(D^0_{\rm phys}\rightarrow l^+X^-)} \; , ~~~~~~~~ 
\bar{r} \; \equiv \; \frac{R(\bar{D}^0_{\rm phys}\rightarrow l^+X^-)}
{R(\bar{D}^0_{\rm phys}\rightarrow l^-X^+)} \; .
\end{equation}
By use of eq. (2.14), we find
\begin{equation}
r \; =\; \left | \frac{q}{p} \right |^2 \frac{1-\alpha}{1+\alpha} \; , ~~~~~~~~
\bar{r} \; =\; \left | \frac{p}{q} \right |^2 \frac{1-\alpha}{1+\alpha} \; , 
\end{equation}
where $\alpha =(1-y^2_D)/(1+x^2_D)$. Note that $|q/p|\neq 1$ signifies 
$CP$ violation in $D^0-\bar{D}^0$ mixing. To fit more accurate data in the near future, 
we prefer the following mixing parameter:
\begin{equation}
r^{~}_D \; \equiv \; \frac{r + \bar{r}}{2} \; =\; w ~
\frac{1-\alpha}{1+\alpha} \; 
\end{equation}
with $w=(|q/p|^2 + |p/q|^2)/2$.
For $|q/p|-1 \sim \pm 1\%$, the value of $w$ deviates less than 
$0.1\%$ from unity. Thus this overall factor of $r^{~}_D$ is safely negligible.
In the approximation of $x^{~}_D\ll 1$ and $y^{~}_D\ll 1$, one obtains
\begin{equation}
r^{~}_D \; \approx \; \frac{x^2_D + y^2_D}{2} \; .
\end{equation}
The latest E691 data \cite{E691} give $r\approx \bar{r} \approx r^{~}_D < 0.37\%$ for small 
$x^{~}_D$ and $y^{~}_D$, where $|q/p| \approx |p/q| \approx 1$, a worse approximation than 
$w \approx 1$, has been used.

\vspace{0.3cm}

The $CP$ asymmetry between a semileptonic decay mode and its $CP$-conjugate counterpart is
defined as
\begin{eqnarray}
\Delta_D & \equiv & \frac{R(\bar{D}^0_{\rm phys}\rightarrow l^+X^-) - 
R(D^0_{\rm phys}\rightarrow l^-X^+)}{R(\bar{D}^0_{\rm phys}\rightarrow l^+X^-)
+ R(D^0_{\rm phys}\rightarrow l^-X^+)} \; , \nonumber \\
\bar{\Delta}_D & \equiv & \frac{R(\bar{D}^0_{\rm phys}\rightarrow l^-X^+) -
R(D^0_{\rm phys}\rightarrow l^+X^-)}{R(\bar{D}^0_{\rm phys}\rightarrow l^-X^+)
+ R(D^0_{\rm phys}\rightarrow l^+X^-)} \; .
\end{eqnarray}
Straightforwardly, we get
\begin{equation}
\Delta_D \; =\; \frac{|p|^4 - |q|^4}{|p|^4 + |q|^4} \; , ~~~~~~~~ \bar{\Delta}_D \; =\; 0 \; .
\end{equation}
If $\Delta_D$ is at the level of $10^{-3}$ or so, it can be measured to three
standard deviations with about $10^{7}$ wrong-sign events. 

\vspace{0.3cm}

It should be noted that
the asymmetry $\bar{\Delta}_D$ may be nonvanishing if there exists new physics affecting
the semileptonic $D$ decays. For example, either the violation of $CPT$ symmetry or
that of the $\Delta Q =\Delta C$ rule can lead to $\bar{\Delta}_D\neq 0$.
Even if the $\Delta Q = \Delta C$ rule and $CPT$ invariance hold, $\bar{\Delta}_D \neq 0$
is still possible in consequence of the phase shifts from final-state electromagnetic
interactions or the $CP$-violating contributions of non-standard electroweak models
to the tree-level processes under discussion. Hence all such fine effects should be
kept in mind and carefully evaluated when one wants to isolate one of them from the others.

\vspace{0.3cm}

As pointed out by Bigi in ref. \cite{Bigi}, a nonvanishing value for $r^{~}_D$ might only be a
secondary signature of $D^0-\bar{D}^0$ mixing, because the presence of $\Delta Q =
-\Delta C$ transitions would contribute to $r^{~}_D$ in a significant and time-independent 
way. For the purpose of illustration, we shall specifically calculate this effect on the 
magnitudes of $r^{~}_D$ and $\bar{\Delta}_D$ in the following.

\begin{center}
{\large\bf B. ~ Effect of $\Delta Q = -\Delta C$ transitions on $r^{~}_D$ and $\bar{\Delta}_D$}
\end{center}

Within the standard model the processes $D^0\rightarrow l^-X^+$ and $\bar{D}^0\rightarrow l^+X^-$
are forbidden according to the $\Delta Q =\Delta C$ rule. New physics beyond the standard model
may allow $\Delta Q =-\Delta C$ transitions, which affect the parameters of $D^0-\bar{D}^0$
mixing and $CP$ violation. In the assumption of $CPT$ symmetry and the neglect of final-state
electromagnetic interactions, the decay amplitudes of
$D^0$ and $\bar{D}^0$ to $l^{\pm}X^{\mp}$ can be factorized as follows:
\begin{eqnarray}
\langle l^+X^-|{\cal H}|D^0\rangle & = & A_l \; , ~~~~~~~~
\langle l^+X^-|{\cal H}|\bar{D}^0\rangle \; =\; \sigma^{~}_l A_l \; ; \nonumber \\
\langle l^-X^+|{\cal H}|\bar{D}^0\rangle & = & A^*_l \; , ~~~~~~~~
\langle l^-X^+|{\cal H}|D^0\rangle \; =\; \sigma^*_l A^*_l \; , 
\end{eqnarray}
where $\sigma^{~}_l$ measures the $\Delta Q =-\Delta C$ transition amplitude. With the
help of eq. (2.14) and notations
\begin{equation}
\lambda_+ \; \equiv \; \frac{q}{p} \sigma^{~}_l \; , ~~~~~~~~~~
\lambda_- \; \equiv \; \frac{p}{q} \sigma^*_l \; ,
\end{equation}
we obtain
\begin{eqnarray}
R(D^0_{\rm phys}\rightarrow l^+X^-) & \propto & |A_l|^2 \left [ (1+\alpha) + (1-\alpha)
|\lambda_+|^2 - 2 y^{~}_D {\rm Re}\lambda_+ - 2\alpha x^{~}_D {\rm Im}\lambda_+ \right ] \; , \nonumber \\
R(\bar{D}^0_{\rm phys}\rightarrow l^-X^+) & \propto & |A_l|^2 \left [ (1+\alpha) +
(1-\alpha) |\lambda_-|^2 - 2y^{~}_D {\rm Re}\lambda_- - 2\alpha x^{~}_D {\rm Im}\lambda_- \right ] \; ;
\end{eqnarray}
and
\begin{eqnarray}
R(\bar{D}^0_{\rm phys}\rightarrow l^+X^-) & \propto & |A_l|^2 \left | \frac{p}{q} \right |^2
\left [(1-\alpha) + (1+\alpha) |\lambda_+|^2 -2 y^{~}_D {\rm Re}\lambda_+ + 
2\alpha x^{~}_D {\rm Im}\lambda_+ \right ] \; , ~~~~ \nonumber \\
R(D^0_{\rm phys}\rightarrow l^-X^+) & \propto & |A_l|^2 \left | \frac{q}{p} \right |^2
\left [ (1-\alpha) + (1+\alpha) |\lambda_-|^2 -2 y^{~}_D {\rm Re}\lambda_- + 
2\alpha x^{~}_D {\rm Im}\lambda_- \right ] \; . ~~~~
\end{eqnarray}
For small $|\sigma^{~}_l|$ (e.g., $|\sigma^{~}_l|\sim x^{~}_D$ or $y^{~}_D$), the
original mixing parameters $r$ and $\bar{r}$ take the following forms:
\begin{eqnarray}
r & \longrightarrow & r^{\prime} \; \approx \; r + |\sigma^{~}_l|^2 
-\frac{2y^{~}_D}{1+\alpha} {\rm Re}\lambda_+ - \frac{2\alpha x^{~}_D}{1+\alpha}
{\rm Im}\lambda_+ \; , \nonumber \\
\bar{r} & \longrightarrow & \bar{r}^{\prime} \; \approx \; \bar{r} + |\sigma^{~}_l|^2
- \frac{2y^{~}_D}{1+\alpha}{\rm Re}\lambda_- - \frac{2\alpha x^{~}_D}{1+\alpha} {\rm Im}\lambda_- \; .
\end{eqnarray}
As a consequence, 
\begin{equation}
r^{\prime}_D \; \equiv \; \frac{r^{\prime}+\bar{r}^{\prime}}{2} \; \approx \; 
r^{~}_D + |\sigma^{~}_l|^2 - \frac{y^{~}_D}{1+\alpha} {\rm Re}(\lambda_+ + \lambda_-)
-\frac{\alpha x^{~}_D}{1+\alpha} {\rm Im}(\lambda_+ + \lambda_-) \; .
\end{equation}
In two extreme cases $\sigma^{~}_l=0$ and $r^{~}_D=0$, we obtain $r^{\prime}_D=r^{~}_D$
and $r^{\prime}_D=|\sigma^{~}_l|^2$ respectively. This implies that a nonzero value for
$r^{\prime}_D$ might not result exclusively from $D^0-\bar{D}^0$ mixing. For this reason,
the study of $D^0-\bar{D}^0$ mixing in some other decay modes of neutral $D$ mesons
(e.g., $D^0/\bar{D}^0\rightarrow K^{\pm}\pi^{\mp}$) is necessary in order to pin down possible
new physics in the charm sector.

\vspace{0.3cm}

The magnitudes of $CP$ asymmetries $\Delta_D$ and $\bar{\Delta}_D$ might be affected
by the $\Delta Q = -\Delta C$ transitions too. In the approximation of $|\sigma^{~}_l|\ll 1$ and
$r^{~}_D\ll 1$, we find that $\bar{\Delta}_D$ becomes
\begin{equation}
\bar{\Delta}^{\prime}_D \; \approx \; r^{~}_D |\sigma^{~}_l|^2 \Delta_D
- \frac{y^{~}_D}{1+\alpha} {\rm Re}( \lambda_- -\lambda_+) - \frac{\alpha x^{~}_D}{1+\alpha}
{\rm Im}(\lambda_- - \lambda_+) \; . 
\end{equation}
Note that nonvanishing $\bar{\Delta}^{\prime}_D$ comes from the interference between the
$D^0-\bar{D}^0$ mixing and $\Delta Q = -\Delta C$ amplitudes, i.e., either $r^{~}_D=0$ or
$\sigma^{~}_l=0$ can give rise to $\bar{\Delta}^{\prime}_D = \bar{\Delta}_D =0$. If 
$\Delta_D =0$ is assumed, then one obtains $\bar{\Delta}^{\prime}_D\approx x^{~}_D 
{\rm Im}\lambda_+$. Since both $x^{~}_D$ and $|\sigma^{~}_l|$ are expected to be very small
(even vanishing), observation of the $CP$ asymmetry $\bar{\Delta}^{\prime}_D$ may be 
practically impossible.

\begin{center}
{\large\bf C. ~ Separate determination of $x^{~}_D$ and $y^{~}_D$}
\end{center}

Current theoretical estimates for the sizes of $x^{~}_D$ and $y^{~}_D$ have dramatical
discrepancies due to the difficulty in dealing with the long-distance interactions
\cite{Wolfenstein-Kaeding,Donoghue-Burdman,Georgi}.
Hence a separate determination of these two mixing parameters from direct measurements
is very necessary \cite{Liu,Xing3}. Here we propose a time-independent method to probe the relative
size of $x^{~}_D$ and $y^{~}_D$ in the dilepton events of coherent $D^0_{\rm phys}
\bar{D}^0_{\rm phys}$ decays at the $\psi(4.14)$ resonance. In our calculations
both $CPT$ invariance and the $\Delta Q=\Delta C$ rule are assumed to hold exactly.

\vspace{0.3cm}

For a $\tau$-charm factory running at the $\psi(4.14)$ resonance, the coherent 
$D^0\bar{D}^0$ events can be produced through $\psi(4.14) \rightarrow \gamma 
(D^0\bar{D}^0)_{C=+}$ or $\psi(4.14)\rightarrow \pi^0 (D^0\bar{D}^0)_{C=-}$, 
where $C$ stands for the charge-conjugation parity \cite{Fry-Ruf}. The generic formulas for the
joint decay rates of two $D$ mesons have been given in eq. (2.21). For our present
purpose, we only consider the primary dilepton events which are directly emitted
from the coherent $(D^0_{\rm phys}\bar{D}^0_{\rm phys})_C$ decays. Let $N^{\pm\pm}_C$
and $N^{+-}_C$ denote the time-integrated numbers of like-sign and opposite-sign
dilepton events, respectively. By use of eq. (2.21), we obtain
\begin{eqnarray}
N^{++}_C & = & N_C \left | \frac{p}{q} \right |^2 \left [ \frac{1+Cy^2_D}{(1-y^2_D)^2}
~ - ~ \frac{1-Cx^2_D}{(1+x^2_D)^2} \right ] \; , \nonumber \\
N^{--}_C & = & N_C \left | \frac{q}{p} \right |^2 \left [ \frac{1+Cy^2_D}{(1-y^2_D)^2}
~ - ~ \frac{1-Cx^2_D}{(1+x^2_D)^2} \right ] \; , \nonumber \\
N^{+-}_C & = & 2 N_C \left [ \frac{1+Cy^2_D}{(1-y^2_D)^2}
~ + ~ \frac{1-Cx^2_D}{(1+x^2_D)^2} \right ] \; ,
\end{eqnarray}
where $N_C$ is the normalization factor proportional to the rates of semileptonic
$D^0$ and $\bar{D}^0$ decays. It is easy to check that the relation
\begin{equation}
N^{++}_- N^{--}_+ \; =\ N^{++}_+ N^{--}_- \;
\end{equation}
holds stringently, and it is independent of the magnitudes of $D^0-\bar{D}^0$ mixing 
and $CP$ violation.

\vspace{0.3cm}

Of course a coherent $D^0\bar{D}^0$ pair with $C=-$ can be straightforwardly produced
from the decay of the $\psi(3.77)$ resonance. Its time-independent decay rates to the
like-sign and opposite-sign dileptons obey eq. (3.14) too. At a 
$\tau$-charm factory the $(D^0\bar{D}^0)_{C=-}$ decays at both the $\psi(3.77)$ and
$\psi(4.14)$ resonances will be measured, and a combination of them might increase
the sensitiveness of our approach to probing $D^0-\bar{D}^0$ mixing.

\vspace{0.3cm}

Usually one is interested in the following two types of observables:
\begin{equation}
a^{~}_C \; \equiv \; \frac{N^{++}_C - N^{--}_C}{N^{++}_C + N^{--}_C} \; , ~~~~~~~~~~
r^{~}_C \; \equiv \; \frac{N^{++}_C + N^{--}_C}{N^{+-}_C} \; ,
\end{equation}
which signify nonvanishing $CP$ violation and $D^0-\bar{D}^0$ mixing, respectively.
Explicitly, we find
\begin{equation}
a^{~}_{-} \; =\; a^{~}_{+} \; =\; \Delta_D \; = \; \frac{|p|^4 - |q|^4}{|p|^4 + |q|^4} \; .
\end{equation}
If $a^{~}_-$ or $a^{~}_+$ is of the order $10^{-3}$,
it can be measured to three standard deviations at the second-round experiments of a 
$\tau$-charm factory with about $10^7$ like-sign dileptons (or equivalently, about 
$10^{10}$ $D^0\bar{D}^0$ events). Furthermore, 
\begin{equation}
r^{~}_{-} \; = \; w ~ \frac{ 1- \alpha}{1+ \alpha} \; , ~~~~~~~~~~
r^{~}_{+} \; = \; w ~ \frac{\beta - \alpha^2}{\beta + \alpha^2} \; ,
\end{equation}
where $\beta = (1+y^2_D)/(1-x^2_D)$. One can see that $r^{~}_- =r^{~}_D$ holds without
any approximation. For small $x^{~}_D$ and $y^{~}_D$, we have 
\begin{equation}
r^{~}_- \; \approx \; \frac{x^2_D +y^2_D}{2} \; ,  ~~~~~~~~
r^{~}_{+} \; \approx \; 3 r^{~}_{-} \; .
\end{equation}
These two approximate results have well been known in the literature (see, e.g., 
refs. \cite{Bigi-Sanda,Bigi}). In such an approximation, however, the 
relative size of $x^2_D$ and $y^2_D$ cannot be determined. 

\vspace{0.3cm}

To distinguish between the different contributions of $x^{~}_D$ and $y^{~}_D$ to 
$D^0-\bar{D}^0$ mixing, one has to measure $r^{~}_{\pm}$ as precisely as possible. 
With the help of eq. (3.18), we show that the magnitudes of $x^{~}_D$ and $y^{~}_D$ 
can be separately determined as follows:
\begin{eqnarray}
x^2_D & = & \left (\frac{1+r^{~}_-}{1-r^{~}_-}\cdot\frac{1+3r^{~}_-}{1-r^{~}_-} 
~ - ~ \frac{1+r^{~}_+}{1-r^{~}_+}\right )
\left (\frac{1+r^{~}_-}{1-r^{~}_-} ~ - ~ \frac{1+r^{~}_+}{1-r^{~}_+} \right )^{-1} \; , \nonumber \\
y^2_D & = & \left (\frac{1-r^{~}_-}{1+r^{~}_-}\cdot\frac{1-3r^{~}_-}{1+r^{~}_-} 
~ - ~ \frac{1-r^{~}_+}{1+r^{~}_+} \right )
\left (\frac{1-r^{~}_+}{1+r^{~}_+} ~ - ~ \frac{1-r^{~}_-}{1+r^{~}_-} \right )^{-1} \; .
\end{eqnarray}
Here it is worth emphasizing that $w$ as the overall (and 
common) factor of $r^{~}_D$, $r^{~}_-$ and $r^{~}_+$ can be safely neglected. In the 
approximations up to $O(r^2_-)$ and $O(r^2_+)$, we obtain two simpler relations:
\begin{equation}
x^2_D - y^2_D \; \approx 2 ~ \frac{r^{~}_+ - 3 r^{~}_-}{r^{~}_+ - r^{~}_-} \; , ~~~~~~~~~~
x^2_D + y^2_D \; \approx \; 4r^{~}_- ~ \frac{r^{~}_+ -2 r^{~}_-}{r^{~}_+ - r^{~}_-} \; .
\end{equation}
Thus it is crucial to examine the deviation of the ratio $r^{~}_+/r^{~}_-$ from 3, in
order to find the difference between $x^2_D$ and $y^2_D$. Instructively, we consider
three special cases \cite{Xing3}:
\begin{eqnarray}
x^{~}_D \; >> \; y^{~}_D  ~~ & \Longrightarrow & ~~ \frac{r^{~}_+}{r^{~}_-} \; \approx \;
3 + 2r^{~}_- \; > \; 3 \; , \nonumber \\
x^{~}_D \; \approx \; y^{~}_D  ~~ & \Longrightarrow  & ~~ \frac{r^{~}_+}{r^{~}_-} \; \approx \;
3 - 9r^{2}_- \; \approx \; 3 \; , \nonumber \\
x^{~}_D \; << \; y^{~}_D  ~~ &  \Longrightarrow  & ~~ \frac{r^{~}_+}{r^{~}_-} \; \approx \;
3 - 2r^{~}_- \; < \; 3 \; .
\end{eqnarray}
These relations can be directly derived from eq. (3.18) or (3.20). If $r^{~}_-$ is
close to the current experimental bound (i.e., $r^{~}_- = r^{~}_D\approx 
(x^2_D+y^2_D)/2 < 0.37\%$), then measurements of $r^{~}_+/r^{~}_-$ to the
accuracy of $10^{-4}$ can definitely establish the relative magnitude of $x^{~}_D$ 
and $y^{~}_D$. To this goal, about $10^8$ like-sign dileptons (or equivalently,
about $10^{11}$ events of $(D^0\bar{D}^0)_{C=-}$ and $(D^0\bar{D}^0)_{C=+}$ pairs) 
are needed. 

\vspace{0.3cm}

For illustration, we take a look at the changes of the measurable
\begin{equation}
\gamma \; \equiv \; \frac{r_+}{r_-} ~ - ~ 3 \; 
\end{equation}
with $x^{~}_D$ by fixing the value of $y^{~}_D$. Allowing 
$10^{-4}\leq r_- < 3.7 \times 10^{-3}$ and taking $y^{~}_D = 0.001$, 0.04 and
0.08 respectively, we plot $\gamma$ as the function of $x^{~}_D$ in fig. 1.
It is clear that $\gamma$ reflects the information about the relative magnitude
of $x^{~}_D$ and $y^{~}_D$, and it can be detected if $r_-$ is of the order
$10^{-3}$ or so.

\vspace{0.3cm}

In the assumption of a dedicated accelerator running for one year at an average luminosity of
$10^{33} {\rm s}^{-1} {\rm cm}^{-2}$, about $10^{7}$ events of $\gamma (D^0\bar{D}^0)
_{C=+}$ and the similar number of $\pi^0 (D^0\bar{D}^0)_{C=-}$ are expected to
be produced at the $\psi(4.14)$ resonance \cite{Fry-Ruf}.
The precision of $10^{-4}$ to $10^{-5}$ in measurements of $r^{~}_-$ and $r^{~}_+$ is 
achievable if one assumes zero background and enough running time \cite{Fry-Ruf,Gladding-Karshon},
and then the similar precision can be obtained for the ratio $r^{~}_+/r^{~}_-$ 
without much more experimental effort (see eq. (3.22) for illustration). 
If $D^0-\bar{D}^0$ mixing were at the level 
of $r^{~}_D\sim 10^{-3}$ (or at least $r^{~}_D\geq 10^{-4}$), then the relative magnitude 
of $x^{~}_D$ and $y^{~}_D$ should be detectable in the second-round experiments 
of a $\tau$-charm factory (beyond the one under consideration at present). 

\section{Neutral $D$ decays to $CP$ eigenstates}
\setcounter{equation}{0}

Neutral $D$-meson decays to hadronic $CP$ eigenstates $f$ (i.e., $|\bar{f}\rangle
\equiv CP|f\rangle = \pm |f\rangle$), such as $f=\pi^+\pi^-$ and $K_S\pi^0$, are of
particular interest for the study of $CP$ violation in the charm sector. The formulas
for their decay rates derived in section 2 can be simplified because of the relations
$\bar{A}_{\bar f}=\pm \bar{A}_f$, $A_{\bar f}=\pm A_f$, $\bar{\rho}^{~}_{\bar f}=1/\rho^{~}_f$
and $\bar{\lambda}_{\bar f}=1/\lambda_f$. If one takes $|q/p|=1$ in some cases, 
then $\bar{\lambda}_{\bar f}=\lambda^*_f$ is obtainable \cite{Du}.

\begin{center}
{\large\bf A. ~ Three sources of $CP$ violation}
\end{center}

In the experimental analyses of incoherent $D$ decays, the combined time-dependent rates
\begin{equation}
{\cal R}_{\pm}(t) \; \equiv \; R(D^0_{\rm phys}(t)\rightarrow f) ~ \pm ~ R(\bar{D}^0_{\rm phys}(t)
\rightarrow f) \; 
\end{equation}
are commonly used. For convenience in expressing our analytical results, we first define
\begin{equation}
{\cal U}_f \; \equiv \; \frac{1-|\lambda_f|^2}{1+|\lambda_f|^2} \; , ~~~~~~~~
{\cal V}_f \; \equiv \; \frac{-2{\rm Im}\lambda_f}{1+|\lambda_f|^2} \; , ~~~~~~~~
{\cal W}_f \; \equiv \; \frac{2{\rm Re}\lambda_f}{1+|\lambda_f|^2} \; ,
\end{equation}
which satisfy a concise sum rule
\begin{equation}
{\cal U}_f^2 + {\cal V}_f^2 + {\cal W}_f^2 \; = \; 1 \; .
\end{equation}
With the help of eq. (2.11), we obtain
\begin{eqnarray}
{\cal R}_+(t) & = & {\cal R}_0 \exp(-\Gamma t) \left [ \cosh(y^{~}_D \Gamma t) - {\cal W}^{~}_f
\sinh(y^{~}_D \Gamma t) \right . \nonumber \\
&  & \left . - \hat{\Delta}_D {\cal U}_f \cos(x^{~}_D \Gamma t) - \hat{\Delta}_D
{\cal V}_f \sin(x^{~}_D \Gamma t) \right ] \; , \nonumber \\
{\cal R}_-(t) & = & {\cal R}_0 \exp(-\Gamma t) \left [ - \hat{\Delta}_D \cosh(y^{~}_D \Gamma t)
+ \hat{\Delta}_D {\cal W}_f \sinh(y^{~}_D \Gamma t) \right . \nonumber \\
&  & \left . + {\cal U}^{~}_f \cos(x^{~}_D \Gamma t) + {\cal V}^{~}_f \sin(x^{~}_D \Gamma t) \right ] \; ,
\end{eqnarray}
where 
\begin{equation}
{\cal R}_0 \; \propto \; \frac{1}{2} |A_f|^2 \left ( 1 + \left | \frac{p}{q} \right |^2 \right )
\left (1+|\lambda_f|^2 \right ) 
\end{equation}
is a normalization factor, and $\hat{\Delta}_D \equiv (|p|^2 - |q|^2)/(|p|^2 + |q|^2)$
is related to $\Delta_D$ through
\begin{equation}
\Delta_D \; = \; \frac{2\hat{\Delta}_D}{1+\hat{\Delta}^2_D} \; .
\end{equation}
To properly describe the signal of direct $CP$ violation in neutral $D$ decays,
we further define
\begin{equation}
{\cal T}_f \; \equiv \; \frac{ 1 ~ - ~ |\rho^{~}_f|^2}{1 ~ + ~ |\rho^{~}_f|^2} \; .
\end{equation}
By use of eq. (2.10), we obtain the relation between ${\cal T}_f$ and ${\cal U}_f$:
\begin{equation}
{\cal U}_f \; =\; \frac{{\cal T}_f ~ + ~ \hat{\Delta}_D}{1 ~ + ~ \hat{\Delta}_D {\cal T}_f} \; .
\end{equation}
It is clear that $\hat{\Delta}_D$, ${\cal T}_f$ and ${\cal V}_f$ measure the $CP$ asymmetry in
$D^0-\bar{D}^0$ mixing, the direct $CP$ asymmetry in the transition amplitudes of $D$ decays,
and the indirect $CP$ asymmetry arising from the interplay of decay and $D^0-\bar{D}^0$ mixing,
respectively. These sources of $CP$-violating effects appear in ${\cal R}_{\pm}(t)$ simultaneously,
but they have different time distributions and can in principle be distinguished from one 
another \cite{Sanda-Xing}. The magnitudes of $\hat{\Delta}_D$, ${\cal T}_f$ and ${\cal V}_f$ 
are expected to be very small (e.g., at the percent level in some extensions of the standard 
electroweak model \cite{Bigi-Gabbiani-Masiero}). In contrast,
the $CP$-conserving quantity ${\cal W}_f$ should be of $O(1)$. Thus the $\cos(x^{~}_D\Gamma t)$
and $\sin(x^{~}_D \Gamma t)$ terms are considerably suppressed in ${\cal R}_+(t)$. This
interesting feature implies that the mixing parameter $y^{~}_D$ is possible to be constrained 
from the measurement of the flavor-untagged decay rate ${\cal R}_+(t)$. We shall discuss this
possibility for some neutral $D$-meson decays in the next subsection.

\vspace{0.3cm}

In lowest-order approximations, we keep only the leading terms of $\hat{\Delta}_D$, ${\cal T}_f$
and ${\cal V}_f$ in ${\cal R}_{\pm}(t)$. Then the $CP$-violating observable is given as
\begin{equation}
{\cal A}(t) \; \equiv \; \frac{{\cal R}_-(t)}{{\cal R}_+(t)} \; \approx \;
-\hat{\Delta}_D ~ + ~ {\cal U}_f ~ + ~ x^{~}_D {\cal V}_f \Gamma t \; \approx \;
{\cal T}_f ~ + ~ x^{~}_D {\cal V}_f \Gamma t \; .
\end{equation}
One can see that $\hat{\Delta}_D$ has little contribution to ${\cal A}(t)$,
and the term ${\cal T}_f$ is almost independent of the decay time $t$.

\vspace{0.3cm}

Integrating ${\cal R}_{\pm}(t)$ over $t\in [0, \infty)$, we obtain the time-independent
decay rates as follows:
\begin{eqnarray}
{\cal R}_+ & = & \frac{{\cal R}_0}{1-y^2_D} \left [ 1 - y^{~}_D {\cal W}_f -
\alpha \hat{\Delta}_D \left ( {\cal U}_f + x^{~}_D {\cal V}_f \right ) \right ] \; , \nonumber \\
{\cal R}_- & = & \frac{{\cal R}_0}{1-y^2_D} \left [ \left (y^{~}_D {\cal W}_f - 1 \right )
\hat{\Delta}_D + \alpha \left ({\cal U}_f + x^{~}_D {\cal V}_f \right ) \right ] \; ,
\end{eqnarray}
where $\alpha$ has been given in section 3-A.
The corresponding $CP$ asymmetry turns out to be
\begin{equation}
{\cal A} \; \equiv \; \frac{{\cal R}_-}{{\cal R}_+} \; \approx \; 
-\hat{\Delta}_D ~ + ~ {\cal U}_f ~ + ~ x^{~}_D {\cal V}_f  \; \approx \; 
{\cal T}_f ~ + ~ x^{~}_D {\cal V}_f \;
\end{equation}
in the leading-order approximation.

\vspace{0.3cm}

At the $\psi(3.77)$ and $\psi(4.14)$ resonances, the produced $D^0\bar{D}^0$ pair may exist
in a coherent state until one of them decays. Hence we can use the semileptonic decay of
one $D$ meson to tag the flavor of the other meson decaying to a flavor-nonspecific $CP$
eigenstate $f$. The time-integrated rates of such joint decays can be read off from
eq. (2.21). We are more interested in the following combinations of decay rates:
\begin{equation}
\Omega_{\pm}(C) \; \equiv \; R(l^-X^+, f)_C ~ \pm ~ R(l^+X^-, f)_C \; .
\end{equation}
After some straightforward calculations, we obtain
\begin{eqnarray}
\Omega_+(C) & = & \Omega_0 \left [ \left (1+Cy^2_D\right ) ~ - ~ \left (1-Cx^2_D\right )
\alpha^2 \hat{\Delta}_D ~ {\cal U}_f \right . \nonumber \\
&  & \left . - ~ (1+C) \left (y^{~}_D {\cal W}_f ~ + ~
x^{~}_D \alpha^2 \hat{\Delta}_D {\cal V}_f \right ) \right ] \; , \nonumber \\
\Omega_-(C) & = & \Omega_0 \left [ -\left (1+Cy^2_D\right ) \hat{\Delta}_D 
~ + ~ \left (1-Cx^2_D\right ) \alpha^2 ~ {\cal U}_f \right . \nonumber \\
&  & \left . + ~ (1+C) \left (y^{~}_D \hat{\Delta}_D {\cal W}_f ~ + ~ x^{~}_D
\alpha^2 {\cal V}_f \right ) \right ] \; , ~~~~~~
\end{eqnarray}
where
\begin{equation}
\Omega_0 \; \propto \; \frac{2{\cal R}_0 |A_l|^2}{\left (1-y^2_D\right )^2} \; ,
\end{equation}
and other quantities have been defined before. Keeping the leading terms of
$\hat{\Delta}_D$, ${\cal T}_f$ and ${\cal V}_f$, we get the $CP$ asymmetries for $C=\pm$ 
cases as follows:
\begin{eqnarray}
{\cal A}_- & \equiv & \frac{\Omega_-(C=-)}{\Omega_+(C=-)} \; \approx \;
-\hat{\Delta}_D ~ + ~ {\cal U}_f \; \approx \; {\cal T}_f \; , \nonumber \\
{\cal A}_+ & \equiv & \frac{\Omega_-(C=+)}{\Omega_+(C=+)} \; \approx \;
-\hat{\Delta}_D ~ + ~ {\cal U}_f ~ + ~ 2x^{~}_D {\cal V}_f \; \approx \;
{\cal T}_f ~ + ~ 2 x^{~}_D {\cal V}_f \; .
\end{eqnarray}
Indeed, ${\cal A}_-$ is exactly independent of the indirect $CP$-violating term
${\cal V}_f$. The asymmetry ${\cal A}_+$ is mainly composed of two sources of $CP$
violation. Comparing eq. (4.15) with eq. (4.11), one can see that there exists an
interesting relation among three time-independent $CP$ measurables:
\begin{equation}
{\cal A}_- ~ + ~ {\cal A}_+ \; \approx \; 2 {\cal A} \; .
\end{equation}
This result should be testable in a variety of neutral $D$ decays to $CP$ eigenstates.

\begin{center}
{\large\bf B. ~ An approach to constrain $y^{~}_D$ and $x^{~}_D$}
\end{center}

It has been pointed out that $y^{~}_D$ might be probed through measurements 
of the singly Cabibbo-suppressed decays $D^0_{\rm phys}(t) \rightarrow K^+K^-$ and $\pi^+\pi^-$, if
$CP$ conservation could hold in them \cite{Liu}. 
This idea can be straightforwardly understood from the
combined decay rates ${\cal R}_+(t)$ in eq. (4.4). Assuming $CP$ invariance, i.e.,
$\hat{\Delta}_D={\cal T}_f={\cal V}_f=0$ and ${\cal W}_f=1$ (or ${\cal W}_f=-1$ for $CP$-odd 
final states), we find
\begin{equation}
{\cal R}_+(t) \; = \; {\cal R}_0 \exp(-\Gamma t) \left [ \cosh(y^{~}_D \Gamma t)
 -  \sinh(y^{~}_D \Gamma t) \right ]
\; =\; {\cal R}_0 \exp\left [-(1+y^{~}_D) \Gamma t\right ] \; 
\end{equation}
with ${\cal R}_0 \propto 2 |A_f|^2$. 
Because $(1+y^{~}_D)\Gamma =\Gamma_{\rm L}$,
the signature of $D^0-\bar{D}^0$ mixing is indeed a deviation of the slope of ${\cal R}_+(t)$
from $\exp(-\Gamma t)$. Since $\Gamma$ can be measured via other approaches, 
one is then able to constrain the magnitude of $y^{~}_D$. The above method depends strongly
upon the assumption of $CP$ conservation in $D$ decays, hence it may not work well in practice.
Subsequently we shall show that a model-independent constraint on $y^{~}_D$ (or $x^{~}_D$) is indeed achievable,
without any special assumption, from measuring the decay-time distributions of 
$D^0_{\rm phys}(t) / \bar{D}^0_{\rm phys}(t) \rightarrow K_{S,L} + \pi^0$ etc.

\vspace{0.3cm}

Exactly speaking, $K_S\pi^0$ and $K_L\pi^0$ are not $CP$ eigenstates due to the existence
of small $CP$ violation in $K^0-\bar{K}^0$ mixing. Here we want to keep this $CP$-violating
contribution to $D$ decays (measured by $\epsilon$), but it can be safely
neglected in most cases.

\vspace{0.3cm}

In the standard model the transitions $D^0\rightarrow \bar{K}^0\pi^0$ and $D^0\rightarrow K^0\pi^0$ (and
their $CP$-conjugate processes) are Cabibbo-allowed and doubly Cabibbo-suppressed, respectively.
Both of them occur only through the tree-level quark diagrams, as illustrated in fig. 2. 
Since any new physics cannot significantly affect the direct decays of charm quark via the
tree-level $W$-mediated graphs \cite{XingDB}, one expects that fig. 2 remains to be a valid quark-diagram
description of the above-mentioned decay modes even beyond the standard model. Indeed significant
new physics may exist in $D^0-\bar{D}^0$ mixing and the loop-induced penguin transitions
of $D$ mesons \cite{Hewett,Blaylock-Seiden-Nir}. The processes $D^0\rightarrow K_{S,L}+\pi^0$ and $\bar{D}^0\rightarrow
K_{S,L}+\pi^0$ take place through fig. 2 with $K^0-\bar{K}^0$ mixing in the final states.
The mass eigenstates of $K^0$ and $\bar{K}^0$ mesons can be written as
\footnote{For simplicity, we neglect the common normalization factor $1/\sqrt{2(1+|\epsilon|^2)}$
for $|K_S\rangle$ and $|K_L\rangle$.}
\begin{eqnarray}
|K_{S}\rangle & = & (1+\epsilon) |K^0\rangle ~ + ~ (1-\epsilon) |\bar{K}^0\rangle  \; , \nonumber \\
|K_{L}\rangle & = & (1+\epsilon) |K^0\rangle ~ - ~ (1-\epsilon) |\bar{K}^0\rangle  \; ,
\end{eqnarray}
where the complex parameter $\epsilon$ has been unambiguously measured ($|\epsilon|\approx
2.27\times 10^{-3}$ and $\phi_{\epsilon}\approx 43.6^0$ \cite{PDG}). Note again that we
do not assume $K_S\pi^0$ and $K_L\pi^0$ to be the exact $CP$ eigenstates, although 
such an assumption is safely allowed by our main results presented later on.
The overall decay amplitudes of $D^0/\bar{D}^0\rightarrow K_{S,L}+\pi^0$ are then given by
\begin{eqnarray}
A(D^0\rightarrow K_{S,L}+\pi^0) & = & 
\left (1+\epsilon^*_{~}\right ) A^{~}_{K^0\pi^0} ~ \pm ~ \left (1-\epsilon^*_{~}\right ) 
A^{~}_{\bar{K}^0_{~}\pi^0_{~}} \; , \nonumber \\
A(\bar{D}^0\rightarrow K_{\rm S,L}+\pi^0) & = & 
\left (1+\epsilon^*\right ) \bar{A}_{K^0\pi^0} ~ \pm ~ \left (1-\epsilon^*\right ) 
\bar{A}_{\bar{K}^0\pi^0} \; .
\end{eqnarray}
Here $A_{K^0\pi^0}$ etc can be factorized as follows:
\begin{eqnarray}
A_{K^0\pi^0} & = & (V_{cd}V^*_{us}) T_1 \exp({\rm i}\delta_1) \; , ~~~~~~~~
A_{\bar{K}^0\pi^0} \; =\; (V_{cs}V^*_{ud}) T_2 \exp({\rm i}\delta_2) \; ; \nonumber \\
\bar{A}_{\bar{K}^0\pi^0} & = & (V^*_{cd}V_{us}) T_1 \exp({\rm i}\delta_1) \; , ~~~~~~~~
\bar{A}_{K^0\pi^0} \; =\; (V^*_{cs}V_{ud}) T_2 \exp({\rm i}\delta_2) \; ,
\end{eqnarray}
where $V_{us}$ etc are the KM matrix elements, $T_1$ and $T_2$ stand for the real 
(positive) hadronic matrix elements, $\delta_1$ and $\delta_2$ are the corresponding strong 
phases. Denoting $h\equiv T_2/T_1$ and $\delta\equiv \delta_2 -\delta_1$, we obtain 
\begin{eqnarray}
\rho^{~}_{K_S\pi^0} & = & + ~ \frac{(1+\epsilon^*) (V^*_{cs}V_{ud}) h \exp({\rm i}\delta)
~ + ~ (1-\epsilon^*) (V^*_{cd}V_{us})}
{(1-\epsilon^*) (V_{cs}V^*_{ud}) h \exp({\rm i}\delta) ~ + ~ (1+\epsilon^*) (V_{cd}V^*_{us})}
\; , \nonumber \\
\rho^{~}_{K_L\pi^0} & = & - ~ \frac{(1+\epsilon^*) (V^*_{cs}V_{ud}) h \exp({\rm i}\delta)
~ - ~ (1-\epsilon^*) (V^*_{cd}V_{us})}
{(1-\epsilon^*) (V_{cs}V^*_{ud}) h \exp({\rm i}\delta) ~ - ~ (1+\epsilon^*) (V_{cd}V^*_{us})} \; .
\end{eqnarray}
By use of the Wolfenstein parameter $\lambda \approx 0.22$, we have $|\epsilon|\approx \lambda^4$,
$V^*_{cs}V_{ud}\approx 1$ and $V^*_{cd}V_{us}\approx -\lambda^2$. Furthermore, 
$h\approx 1$ is anticipated in the factorization approximation \cite{Bigi-Yamamoto}. As a consequence,
\begin{equation}
\rho^{~}_{K_S\pi^0} \; \approx \; 1 + 2 \epsilon^* \; , ~~~~~~~~
\rho^{~}_{K_L\pi^0} \; \approx \; - \rho^{~}_{K_S\pi^0} \; 
\end{equation}
hold to a good degree of accuracy. This result implies that the direct $CP$ asymmetries
in $D^0_{\rm phys}(t)/\bar{D}^0_{\rm phys}(t)\rightarrow K_{S,L}+\pi^0$ are dominated by
$K^0-\bar{K}^0$ mixing \cite{Xing4}:
\begin{equation}
{\cal T}_{K_S\pi^0} \; \approx \; {\cal T}_{K_L\pi^0} \; \approx \;
-2 {\rm Re}\epsilon \; \approx \; -2 |\epsilon| \cos\phi_{\epsilon} \; .
\end{equation}
Explicitly, we get ${\cal T}_{K_S\pi^0} \approx {\cal T}_{K_L\pi^0}\approx 
-3.3\times 10^{-3}$. 

\vspace{0.3cm}

For simplicity, we shall use the notation $q/p = |q/p| \exp({\rm i}2\phi)$ (see eq. (2.2))
later on. With the help of eqs. (4.22) and (4.23) as well as the reasonable assumption 
$|\hat{\Delta}_D|\leq 10^{-2}$, we obtain
\begin{eqnarray}
{\cal U}_{K_S\pi^0} & \approx & + ~ {\cal U}_{K_L\pi^0} \; \approx \;
\hat{\Delta}_D ~ - ~ 2 |\epsilon| \cos\phi_{\epsilon} \; , \nonumber \\
{\cal V}_{K_S\pi^0} & \approx & - ~ {\cal V}_{K_L\pi^0} \; \approx \;
2 |\epsilon| \sin\phi_{\epsilon} \cos (2\phi) ~ - ~ \sin(2\phi) \; , \nonumber \\
{\cal W}_{K_S\pi^0} & \approx & - ~ {\cal W}_{K_L\pi^0} \; \approx \;
2 |\epsilon| \sin\phi_{\epsilon} \sin(2\phi) ~ + ~ \cos(2\phi) \;
\end{eqnarray}
in good approximations. Clearly the unknown new physics may enter
${\cal V}_{K_S\pi^0}$ and ${\cal W}_{K_S\pi^0}$ through the $D^0-\bar{D}^0$ mixing
phase $\phi$. Within the standard model one expects $\phi\sim 0$, thus ${\cal V}_{K_S\pi^0}
\approx 3.1\times 10^{-3}$ and ${\cal W}_{K_S\pi^0}\approx 1$. Beyond the standard model
it is possible that the magnitudes of ${\cal V}_{K_S\pi^0}$ and ${\cal W}_{K_S\pi^0}$ are dominated by
$\sin(2\phi)$ and $\cos(2\phi)$, respectively. 
The quantities ${\cal V}_{K_L\pi^0}$ and ${\cal W}_{K_L\pi^0}$ are in the similar situation.

\vspace{0.3cm}

Due to the smallness of $x^{~}_D$ and $y^{~}_D$, some analytical approximations can be made
for ${\cal R}_{\pm}(t)$ in. eq. (4.4) up to $O(x^2_D)$ and $O(y^2_D)$. Taking eqs. (4.23) and (4.24) into
account, we find
\begin{eqnarray}
{\cal R}^{K_S\pi^0}_{+}(t) & \approx & {\cal R}^{K_S\pi^0}_0 
\exp (-\Gamma t) \left [ 1 ~ + ~ X \Gamma^2 t^2 ~ - ~ Y \Gamma t \right ] \; , \nonumber \\
{\cal R}^{K_L\pi^0}_{+}(t) & \approx & {\cal R}^{K_L\pi^0}_0 
\exp (-\Gamma t) \left [ 1 ~ + ~ X \Gamma^2 t^2 ~ + ~ Y \Gamma t \right ] \; , 
\end{eqnarray}
where $X$ and $Y$ are functions of $x^{~}_D$ and $y^{~}_D$:
\begin{eqnarray}
X & \approx & \frac{1}{2} \left [ y^2_D ~ + ~ x^2_D \hat{\Delta}_D \left (
\hat{\Delta}_D - 2 |\epsilon| \cos\phi_{\epsilon} \right ) \right ] \; , \nonumber \\
Y & \approx & 2 |\epsilon| \sin\phi_{\epsilon} \left [ y^{~}_D \sin(2\phi)
+ x^{~}_D \hat{\Delta}_D \cos(2\phi) \right ] 
~ + ~ y^{~}_D \cos(2\phi) ~ - ~ x^{~}_D \hat{\Delta}_D \sin(2\phi) \; .
\end{eqnarray}
We can see that $X$ and $Y$ vanish in the absence of $D^0-\bar{D}^0$ mixing, and the contribution
of $x^{~}_D$ to them is significantly suppressed by $\hat{\Delta}_D$. 
Naively one might expect to measure the deviations
of ${\cal R}^{K_S\pi^0}_+(t)$ and ${\cal R}^{K_L\pi^0}_+(t)$ from 
${\cal R}^{K_S\pi^0}_0\exp(-\Gamma t)$ and ${\cal R}^{K_L\pi^0}_0\exp(-\Gamma t)$, 
respectively, in order to determine the sizes of $X$ and $Y$. However this is very 
difficult, if not even practically impossible, because of the smallness of $X$ and $Y$.
The interesting point here is that a comparison between the time distributions of
${\cal R}^{K_S\pi^0}_+(t)$ and ${\cal R}^{K_L\pi^0}_+(t)$ can definitely
constrain the magnitude of $Y$. In view of $|\epsilon|\sim 10^{-3}$, $x^{~}_D<0.086$,
$y^{~}_D<0.086$ and $|\hat{\Delta}_D|\leq 10^{-2}$ from our present experimental 
knowledge (and theoretical expectation), only the $y^{~}_D\cos(2\phi)$ term of $Y$
is possible to be at the percent level (magnitudes of the other three terms in $Y$
are all below $10^{-3}$). One can conclude that the detectable
signal of $Y$ has to be at the percent level and it must come mainly from the width
difference of $D^0$ and $\bar{D}^0$ mass eigenstates. For illustration, the time distributions of 
${\cal R}^{K_S\pi^0}_+(t)$ and ${\cal R}^{K_L\pi^0}_+(t)$ are depicted in
fig. 3 by taking $y^{~}_D\approx 0.08$ and $\phi\approx 0$. We see that
around $\Gamma t=2$ the difference between ${\cal R}^{K_S\pi^0}_+(t)/{\cal R}^{K_S\pi^0}_0$
and ${\cal R}^{K_L\pi^0}_+(t)/{\cal R}^{K_L\pi^0}_0$ can be as large as $5\%$,
allowing us to extract a signal of $D^0-\bar{D}^0$ mixing provided that the accuracy
of practical measurements is good enough.

\vspace{0.3cm}

The asymmetry between ${\cal R}^{K_S\pi^0}_+(t)$ and ${\cal R}^{K_L\pi^0}_+(t)$ can be
given as
\begin{equation}
{\cal A}_{LS}(t) \; \equiv \; \frac{{\cal R}^{K_S\pi^0}_+(t) - {\cal R}^{K_L\pi^0}_+(t)}
{{\cal R}^{K_S\pi^0}_+(t) + {\cal R}^{K_L\pi^0}_+(t)} 
\; \approx \; -2\lambda^2 \frac{\cos\delta}{h} \left ( 1 + X \Gamma^2 t^2 \right )
 ~ - ~ Y \Gamma t \; .
\end{equation}
Indeed the coefficient $-2\lambda^2 \cos\delta/h$ measures the decay-rate asymmetry between
$D^0\rightarrow K_S + \pi^0$ and $D^0\rightarrow K_L + \pi^0$ (or their flavor-conjugate
processes) \cite{Sanda-Xing,Bigi-Yamamoto}. The measurement of ${\cal A}_{LS}(t)$ allows us 
to extract the magnitude of $Y$. To give one a numerical feeling, the changes of 
${\cal A}_{LS}(t)$ with $t$ are illustrated in fig. 4 by assuming $X\approx y^2_D/2$,
$Y\approx y^{~}_D \cos(2\phi)$, $h\approx 1$, $\delta\approx 0$ and taking
$y^{~}_D \approx 0.08$, $|\cos(2\phi)|\approx 1$. It is clear that a large
signal of $y^{~}_D$ should be detectable from ${\cal A}_{LS}(t)$.

\vspace{0.3cm}

The effects of $D^0-\bar{D}^0$ mixing and $CP$ violation also manifest themselves in
the combined rates ${\cal R}^{K_S\pi^0}_-(t)$ and ${\cal R}^{K_L\pi^0}_-(t)$:
\begin{eqnarray}
{\cal R}^{K_S\pi^0}_{-}(t) & \approx & {\cal R}^{K_S\pi^0}_0 
\exp (-\Gamma t) \left [ - 2|\epsilon| \cos\phi_{\epsilon} ~ + ~ X^{\prime} \Gamma^2 t^2 ~ + ~ 
Y^{\prime} \Gamma t \right ] \; , \nonumber \\
{\cal R}^{K_L\pi^0}_{-}(t) & \approx & {\cal R}^{K_L\pi^0}_0 
\exp (-\Gamma t) \left [ - 2|\epsilon| \cos\phi_{\epsilon} ~ + ~ X^{\prime} \Gamma^2 t^2 ~ - ~ 
Y^{\prime} \Gamma t \right ] \; , 
\end{eqnarray}
where 
\begin{eqnarray}
X^{\prime} & \approx & x^2_D |\epsilon| \cos\phi_{\epsilon} ~ - ~ r^{~}_D \hat{\Delta}_D \; , \nonumber \\
Y^{\prime} & \approx & 2 |\epsilon| \sin\phi_{\epsilon} \left [ y^{~}_D \hat{\Delta}_D \sin(2\phi)
+ x^{~}_D \cos(2\phi) \right ] ~ + ~ y^{~}_D \hat{\Delta}_D \cos(2\phi) ~ - ~ x^{~}_D \sin(2\phi) \; 
\end{eqnarray}
with $r^{~}_D\approx (x^2_D+y^2_D)/2$. Obviously the $CP$ asymmetry induced by $K^0-\bar{K}^0$
mixing (i.e., ${\rm Re}\epsilon$) plays an important role in the decay modes under discussion.
The contribution of $CP$ violation in $D^0-\bar{D}^0$ mixing (i.e., $\hat{\Delta}_D$) to $Y^{\prime}$
is not significant even if $\phi\sim 0$. If new physics considerably enhances $x^{~}_D$
and $\phi$, e.g., $x^{~}_D\sim 10^{-2}$ and $|\sin(2\phi)|\sim 1$, then
${\cal R}^{K_S\pi^0}_-(t)$ and ${\cal R}^{K_L\pi^0}_-(t)$ will be dominated by the $CP$
asymmetry arising from the interplay of decay and $D^0-\bar{D}^0$ mixing. In other words,
the signals of $CP$ asymmetries
\begin{eqnarray}
{\cal A}_{K_S\pi^0}(t) & \equiv & \frac{{\cal R}^{K_S\pi^0}_-(t)}
{{\cal R}^{K_S\pi^0}_+(t)} \; \approx \;
-2 |\epsilon| \cos\phi_{\epsilon} ~ + ~ X^{\prime} \Gamma^2 t^2 ~ + ~ Y^{\prime} \Gamma t \; , \nonumber \\
{\cal A}_{K_L\pi^0}(t) & \equiv & \frac{{\cal R}^{K_L\pi^0}_-(t)}
{{\cal R}^{K_L\pi^0}_+(t)} \; \approx \;
-2 |\epsilon| \cos\phi_{\epsilon} ~ + ~ X^{\prime} \Gamma^2 t^2 ~ - ~ Y^{\prime} \Gamma t \; 
\end{eqnarray}
at the percent level will indicate that new physics is definitely present in $D^0-\bar{D}^0$
mixing (e.g., $|\sin(2\phi)|\geq 0.5$) and the magnitude of $x^{~}_D$ must be of 
order $10^{-2}$. Taking $x^{~}_D\approx 0.08$, $\phi\approx \pi/4$ and
$\hat{\Delta}_D\approx 0$ for example, we illustrate the time distributions of ${\cal R}^{K_S\pi^0}_-(t)$
and ${\cal R}^{K_L\pi^0}_-(t)$ in fig. 5. We find that around $\Gamma t=1$ the magnitudes
of the decay-rate differences between $D^0_{\rm phys}(t)\rightarrow K_{S,L}+\pi^0$ and
$\bar{D}^0_{\rm phys}(t)\rightarrow K_{S,L}+\pi^0$ can be as large as $3\%$. Since 
${\cal R}^{K_S\pi^0}_-(1/\Gamma)\sim - {\cal R}^{K_L\pi^0}_-(1/\Gamma)$, it is 
possible to extract the rough size of $Y^{\prime}\approx -x^{~}_D \sin(2\phi)$.
Clearly the measurements of ${\cal R}^{K_S\pi^0}_+(t)$, ${\cal R}^{K_L\pi^0}_+(t)$ and
${\cal R}^{K_S\pi^0}_-(t)$, ${\cal R}^{K_L\pi^0}_-(t)$ are complementary to one
another and can shed some light on the mixing parameters $x^{~}_D$ and $y^{~}_D$ as well as
the possible new physics hidden in $D^0-\bar{D}^0$ mixing.

\vspace{0.3cm}

Note that the above discussions can be directly extended to neutral $D$ decays
to the final states like $K_{S,L}+\rho^0$, $K_{S,L}+a^0_1$ and $K_{S,L}+\omega$,
which occur through the same quark diagrams as $D^0/\bar{D}^0\rightarrow K_{S,L}+\pi^0$
(see fig. 2). Because $X^{(\prime)}$ and $Y^{(\prime)}$ depend only upon the $D^0-\bar{D}^0$
and $K^0-\bar{K}^0$ mixing parameters, a sum over the above modes is possible, without any
dilution effect on the signals of $D^0-\bar{D}^0$ mixing and $CP$ violation, to increase the
number of decay events in statistics.

\begin{center}
{\large\bf C. ~ Final-state interactions in $D\rightarrow K\bar{K}$}
\end{center}

Recently the CLEO Collaboration has searched for $CP$ violation in neutral $D$ decays to
the $CP$ eigenstates $K^+K^-$, $K_S \phi$ and $K_S \pi^0$. 
The confidence intervals ($90\%$) on $CP$ asymmetries in these three modes were 
found to be $-0.020 < {\cal A}_{K\bar{K}} < 0.180$, $-0.182 < {\cal A}_{K_S\phi} <
0.126$ and $-0.067 < {\cal A}_{K_S \pi^0} < 0.031$, respectively \cite{E687-CLEO}. Although
a definite signal of $CP$ violation was not established from the data above, 
the possibility that these
decays may accommodate $CP$ asymmetries at the percent level could not be ruled out.
In the following we shall concentrate on the final-state interactions 
in $D^0/\bar{D}^0\rightarrow K^+K^-$ and $K^0\bar{K}^0$, since they may affect the 
magnitudes of $CP$ asymmetries significantly. The similar
discussions can be extended to some other decay modes such as 
$D^0/\bar{D}^0\rightarrow \pi^+\pi^-$ and $K_S \pi^0$.

\vspace{0.3cm}

We begin with an isospin analysis of $D^0\rightarrow K^+K^-$, $D^+\rightarrow K^+\bar{K}^0$ and
$D^0\rightarrow K^0\bar{K}^0$. To do this, we assume that there is no mixture of
$D\rightarrow K\bar{K}$ with other channels.
In the language of quark diagrams \cite{XingQD}, 
these modes can occur through both tree-level and penguin diagrams.
However, such a naive description is problematic due to the presence of final-state
rescattering effects \cite{Pham-Kamal}. The final states $K\bar{K}$ 
may contain $I=1$ and $I=0$ isospin configurations, and the overall decay amplitudes of 
$D\rightarrow K\bar{K}$ can be written as 
\begin{eqnarray}
A_{+-} & \equiv & \langle K^+ K^- |{\cal H}|D^0\rangle \; = \;
\frac{1}{2} \left ( A_1 ~ + ~ A_0 \right ) \; , \nonumber \\
A_{00} & \equiv & \langle K^0\bar{K}^0 |{\cal H}|D^0\rangle \; = \;
\frac{1}{2} \left ( A_1 ~ - ~ A_0 \right ) \; , \nonumber \\
A_{+0} & \equiv & \langle K^+ \bar{K}^0 |{\cal H}|D^+\rangle \; =\;
A_1 \; ,
\end{eqnarray}
where $A_1$ and $A_0$ are two isospin amplitudes. 
Clearly three decay amplitudes can form an isospin triangle in the complex plane:
$A_{+-} + A_{00} = A_{+0}$. Since the branching ratios of $D\rightarrow K\bar{K}$
have been measured, one is able to determine
the relevant isospin amplitudes from the relations above. For our purpose, we are more
interested in the ratio of two isospin amplitudes: $A_0/A_1 \equiv Z \exp({\rm i}\varphi)$.
It is straightforward to obtain
\begin{equation}
Z \; =\; \sqrt{2 \left (R_{+-} + R_{00} \right ) - 1} \; ,
~~~~~~~~
\cos\varphi \; =\; \frac{R_{+-} - R_{00}}{Z} \; ,
\end{equation}
where $R_{+-} \equiv |A_{+-}/A_{+0}|^{2}$ and $R_{00}=|A_{00}/A_{+0}|^{2}$ are 
two observables. If the annihilation diagrams and penguin effects in $D\rightarrow K\bar{K}$ are negligible, 
then $A_{+-}$, $A_{00}$ and $A_{+0}$ have a common KM factor (i.e., $V_{cs}V^*_{us}$)
from the dominant tree-level (spectator) quark transitions. 
In this case, $\varphi$ is purely a strong phase shift and the magnitude of $Z$ is 
independent of the KM matrix elements.

\vspace{0.3cm}

Current experimental data give $B(D^0\rightarrow K^+K^-) = (4.54 \pm 0.29) \times 10^{-3}$,
$B(D^0\rightarrow K^0\bar{K}^0) = (1.1 \pm 0.4) \times 10^{-3}$ and $B(D^+\rightarrow K^+\bar{K}^0)
= (7.8 \pm 1.7) \times 10^{-3}$ \cite{PDG}. The lifetimes of $D^0$ and $D^+$ mesons are 
$\tau^{~}_{D^0} = (0.415 \pm 0.004) \times 10^{-12}$s and $\tau^{~}_{D^+} = (1.057 \pm 0.015)
\times 10^{-12}$s, respectively. In the neglect of small phase space differences of three decay modes,
we obtain $R_{+-} = 1.48 \pm 0.45$ and $R_{00} = 0.36 \pm 0.22$. The sizes of 
$Z$ and $\varphi$ can be solved from eq. (4.32) with the inputs of $R_{+-}$ and $R_{00}$, but there is
large error propagation in this procedure, particularly for $\cos\varphi$ which
is bounded by unity. For simplicity and illustration, 
we plot the allowed regions of $Z$ and $\cos\varphi$ in fig. 6.
One can observe that $1.7 \leq Z \leq 2.0$ and $0.3 \leq \cos\varphi \leq 1.0$ (the central
values of $R_{+-}$ and $R_{00}$ lead to $Z\approx 1.6$ and $\cos\varphi \approx 0.68$). 
This implies that significant final-state interactions may exist in the processes $D\rightarrow K\bar{K}$.

\vspace{0.3cm}

The isospin amplitudes $A_1$ and $A_0$ can be expanded in terms of the tree-level and penguin
transition amplitudes \cite{Palmer-Wu}. Without loss of generality, we write
\footnote{Here we have neglected the contributions of tree-level annihilation
diagrams to $D^0\rightarrow K^0\bar{K}^0$, which involve both $V_{cs}V^*_{us}$
and $V_{cd}V^*_{ud}$. These two graph amplitudes are expected to have large cancellation
with each other due to the GIM mechanism \cite{Pham-Kamal}.}
\begin{eqnarray}
A_1 & = & A_{1 \rm T} \exp\left [{\rm i} \left (\phi_{\rm T} + \delta_{1 \rm T} \right ) \right ]
~ + ~ A_{1 \rm P} \exp\left [{\rm i} \left (\phi_{\rm P} + \delta_{1 \rm P} \right ) \right ] \; ,
\nonumber \\
A_0 & = & A_{0 \rm T} \exp\left [{\rm i} \left (\phi_{\rm T} + \delta_{0 \rm T} \right ) \right ]
~ + ~ A_{0 \rm P} \exp\left [{\rm i} \left (\phi_{\rm P} + \delta_{0 \rm P} \right ) \right ] \; ,
\end{eqnarray}
where $\phi_{\rm T}$ and $\phi_{\rm P}$ are the overall weak phases of tree-level
and penguin diagrams respectively,
$\delta_{\rm n T}$ and $\delta_{\rm n P}$ (with n=1,0)
denote the corresponding strong phases. Hence $Z$ and $\varphi$ defined above are 
complicated functions of $A_{\rm n T}$, $A_{\rm n P}$, $\phi_{\rm T}$, $\phi_{\rm P}$,
$\delta_{\rm n T}$ and $\delta_{\rm n P}$. Since new physics may significantly affect the
penguin amplitudes, direct $CP$ violation is possible to appear in $D\rightarrow K\bar{K}$.
A constraint on the $I=1$ penguin contribution to $D\rightarrow K\bar{K}$ can be obtained by
observing the decay-rate asymmetry between $D^+\rightarrow K^+\bar{K}^0$ and
$D^-\rightarrow K^-K^0$:
\begin{eqnarray}
&  & \frac{|\langle K^+\bar{K}^0|{\cal H}|D^+\rangle |^2 - |\langle K^-K^0|{\cal H}|D^-\rangle |^2}
{|\langle K^+\bar{K}^0|{\cal H}|D^+\rangle |^2 + |\langle K^-K^0|{\cal H}|D^-\rangle |^2} \nonumber \\
\nonumber \\
& = & \frac{-2 A_{1 \rm T} A_{1 \rm P} \sin (\phi_{\rm P} - \phi_{\rm T} ) 
\sin (\delta_{1 \rm P} - \delta_{1 \rm T} )}
{A^2_{1 \rm T} + A^2_{1 \rm P} + 2 A_{1 \rm T} A_{1 \rm P} \cos (\phi_{\rm P} - \phi_{\rm T} )
\cos ( \delta_{1 \rm P} - \delta_{1 \rm T} )} \; .
\end{eqnarray}
Note that the weak phase difference $|\phi_{\rm P} - \phi_{\rm T}|$ may be rather
small within the standard model, but some sources of new physics (e.g., the existence 
of the fourth quark family or an iso-singlet up-type quark \cite{Nir-Branco}) can
significantly enhance it through the breakdown of unitarity of the $3\times 3$ KM
matrix in the penguin loops \cite{XingNPB}.
The direct $CP$ asymmetries in $D^0/\bar{D}^0\rightarrow K^+K^-$ and $K^0\bar{K}^0$ 
contain both $I=1$ and $I=0$ penguin contributions, and the latter can in principle
be distinguished from the former with the help of eq. (4.34). In practical experiments,
${\cal T}_{K^+K^-}$ and ${\cal T}_{K^0\bar{K}^0}$ are cleanly detectable on the 
$\psi(3.77)$ resonance (see eq. (4.15) for illustration). If one wants to calculate
the decay-rate asymmetries between $D^{\pm}\rightarrow K^{\pm} + K_{S,L}$ or
between $D^0/\bar{D}^0\rightarrow K_{S,L} + K_{S,L}$, then the $CP$ violation induced
by $K^0-\bar{K}^0$ mixing in the final states has to be taken into account.

\vspace{0.3cm}

It is also argued that inelastic final-state interactions may affect $D\rightarrow 
K\bar{K}$ \cite{Pham-Kamal}. This kind of effect is possible to yield 
a nonvanishing rate asymmetry between the charged $D$ decays to $K^+\bar{K}^0$
and $K^-K^0$, even though the penguin contributions are negligibly small.
To justify the role of penguin transitions and inelastic
final-state interactions, one has to rely on
the future data on direct $CP$ asymmetries in the decay modes under discussion.

\section{Neutral $D$ decays to non-$CP$ eigenstates}
\setcounter{equation}{0}

We proceed to consider the case that both $D^0$ and $\bar{D}^0$ mesons decay to a 
common non-$CP$ eigenstate. Most of such decay modes occur through quark 
transitions of the types $c\rightarrow s(u\bar{d})$ and $c\rightarrow d(u\bar{s})$
or their flavor-conjugate counterparts, and the typical examples are the
Cabibbo-allowed decay $D^0\rightarrow K^-\pi^+$ and the doubly Cabibbo-suppressed
process $D^0\rightarrow K^+\pi^-$. Because neutral $D$ decays to $K^{\pm}\pi^{\mp}$
are of particular interest for the study of $D^0-\bar{D}^0$ mixing and DCSDs in
charm physics, we shall concentrate on them in this section. Of course similar
discussions can be extended to other non-$CP$ eigenstates. 

\vspace{0.3cm}

Note that $D^0\rightarrow K^{\pm}\pi^{\mp}$ and their $CP$-conjugate processes take 
place only via the tree-level quark diagrams, on which no new physics can have 
significant effect \cite{Xing2,Blaylock-Seiden-Nir}. 
Thus the four transition amplitudes are factorized as follows:
\begin{eqnarray}
A_{K^-\pi^+} & = & (V_{cs}V^*_{ud}) T_a \exp({\rm i}\delta_a) \; , ~~~~~~~~
A_{K^+\pi^-} \; =\; (V_{cd}V^*_{us}) T_b \exp({\rm i}\delta_b) \; , \nonumber \\
\bar{A}_{K^+\pi^-} & = & (V^*_{cs}V_{ud}) T_a \exp({\rm i}\delta_a) \; , ~~~~~~~~
\bar{A}_{K^-\pi^+} \; =\; (V^*_{cd}V_{us}) T_b \exp({\rm i}\delta_b) \; , 
\end{eqnarray}
where $T_a$ and $T_b$ denote the real (positive) hadronic matrix elements, 
$\delta_a$ and $\delta_b$ are the corresponding strong phases. Defining 
$h_{K\pi}\equiv T_b/T_a$ and $\delta_{K\pi}\equiv \delta_b -\delta_a$, we obtain
\begin{equation}
\rho^{~}_{K^-\pi^+} \; \approx \; \bar{\rho}^{~}_{K^+\pi^-} \; \approx \;
-\lambda^2 h_{K\pi} \exp({\rm i}\delta_{K\pi}) \; 
\end{equation}
to a good degree of accuracy, where $\lambda\approx 0.22$ is the Wolfenstein
parameter. In the factorization approximation, the magnitude of $h_{K\pi}$ is
expected to be of $O(1)$. The strong phase shift $\delta_{K\pi}$ vanishes only in
the limit of $SU(3)$ symmetry \cite{Wolfenstein}. To fit the recent CLEO result
for $D^0\rightarrow K^{\pm}\pi^{\mp}$ \cite{CLEO}, which gives $|\rho^{~}_{K^-\pi^+}|^2
=(0.77\pm 0.25\pm 0.25)\%$, one finds $\delta_{K\pi}\sim 5^0 - 13^0$ from a 
few phenomenological models \cite{Chau-Cheng,Buccella,Browder-Pakvasa}.
Of course a larger value for $\delta_{K\pi}$ cannot be absolutely ruled out from current
experimental data because of many uncertainties associated with the empiric
models used to analyze nonleptonic $D$ decays.
Finally, the expressions of $\lambda_{K^-\pi^+}$ and $\bar{\lambda}_{K^+\pi^-}$
are obtainable from eq. (5.2) as 
\begin{eqnarray}
\lambda_{K^-\pi^+} & \approx & -\lambda^2 h_{K\pi} \left | \frac{q}{p} \right |
\exp\left [{\rm i}(\delta_{K\pi} +2\phi) \right ] \; , \nonumber \\
\bar{\lambda}_{K^+\pi^-} & \approx & -\lambda^2 h_{K\pi} \left | \frac{p}{q} \right |
\exp\left [{\rm i}(\delta_{K\pi} -2\phi)\right ] \; ,
\end{eqnarray}
where we have used the notation of $q/p$ given in eq. (2.2).

\begin{center}
{\large\bf A. ~ Incoherent $D$ decays to $K^{\pm}\pi^{\mp}$}
\end{center}

A lot of attention has been paid to the time distributions of incoherent $D$
decays to $K^{\pm}\pi^{\mp}$ (see, e.g., refs. \cite{Liu,Blaylock-Seiden-Nir,Browder-Pakvasa}).
In particular, Browder and Pakvasa have given a
quite detailed analysis of the implications of $CP$ violation and final-state
interactions in the search for $D^0-\bar{D}^0$ mixing from $D^0_{\rm phys}(t)
\rightarrow K^+\pi^-$ and $\bar{D}^0_{\rm phys}(t)\rightarrow K^-\pi^+$
\cite{Browder-Pakvasa}. Our subsequent discussions are complementary to
their work on three points: (a) the $CP$-violating asymmetry between
$D^0_{\rm phys}(t)\rightarrow K^-\pi^+$ and $\bar{D}^0_{\rm phys}(t)\rightarrow
K^+\pi^-$ is analyzed; (b) the different effects of $x^{~}_D$ and $y^{~}_D$ 
on $D^0_{\rm phys}(t)\rightarrow K^{\pm}\pi^{\mp}$ and $\bar{D}^0_{\rm phys}(t)
\rightarrow K^{\pm}\pi^{\mp}$ are explored in detail; and (c) the time-independent
measurements of these decay modes are considered.

\vspace{0.3cm}

Up to $O(x^2_D)$, $O(y^2_D)$ and $O(\lambda^4)$ for every distinctive term, 
the decay rates of $D$ to $K^{\pm}\pi^{\mp}$ can be directly read off from
eqs. (2.22) and (2.23):
\begin{eqnarray}
R(D^0_{\rm phys}(t)\rightarrow K^-\pi^+) & \propto & |A_{K^-\pi^+}|^2 \exp(-\Gamma t) \left \{
1 + \lambda^2 h_{K\pi} \left | \frac{q}{p} \right | \left [ y^{~}_D \cos\left (\delta^{~}_{K\pi}+
2\phi\right ) \right . \right . \nonumber \\
&  & \left . \left . + x^{~}_D \sin\left (\delta^{~}_{K\pi} +2\phi\right ) \right ]
\Gamma t - \frac{1}{4} \left (x^2_D -y^2_D\right ) \Gamma^2 t^2 \right \} \; , \nonumber \\
R(\bar{D}^0_{\rm phys}(t)\rightarrow K^+\pi^-) & \propto & |A_{K^-\pi^+}|^2 \exp(-\Gamma t) \left \{
1 + \lambda^2 h_{K\pi} \left | \frac{p}{q} \right | \left [ y^{~}_D \cos\left (\delta^{~}_{K\pi}-
2\phi\right ) \right . \right . \nonumber \\
&  & \left . \left . + x^{~}_D \sin\left (\delta^{~}_{K\pi} -2\phi\right ) \right ]
\Gamma t - \frac{1}{4} \left (x^2_D -y^2_D\right ) \Gamma^2 t^2 \right \} \; ;
\end{eqnarray}
and
\begin{eqnarray}
R(D^0_{\rm phys}(t)\rightarrow K^+\pi^-) & \propto & |A_{K^-\pi^+}|^2 \exp(-\Gamma t) \left \{
\lambda^4 h^2_{K\pi} + \lambda^2 h_{K\pi} \left | \frac{q}{p} \right | \left [ y^{~}_D \cos\left (\delta^{~}_{K\pi}- 2\phi\right ) \right . \right . \nonumber \\
&  & \left . \left . - x^{~}_D \sin\left (\delta^{~}_{K\pi} -2\phi\right ) \right ]
\Gamma t + \frac{r}{2} \Gamma^2 t^2 \right \} \; , \nonumber \\
R(\bar{D}^0_{\rm phys}(t)\rightarrow K^-\pi^+) & \propto & |A_{K^-\pi^+}|^2 \exp(-\Gamma t) \left \{
\lambda^4 h^2_{K\pi} + \lambda^2 h_{K\pi} \left | \frac{p}{q} \right | \left [ y^{~}_D \cos\left (\delta^{~}_{K\pi}+ 2\phi\right ) \right . \right . \nonumber \\
&  & \left . \left . - x^{~}_D \sin\left (\delta^{~}_{K\pi} +2\phi\right ) \right ]
\Gamma t + \frac{\bar{r}}{2} \Gamma^2 t^2 \right \} \; ,
\end{eqnarray}
where $r$ and $\bar{r}$ have been presented in eq. (3.2). To probe $CP$ violation and
$D^0-\bar{D}^0$ mixing, the following two types of measurables can be analyzed in experiments:

\vspace{0.3cm}

(1) The $CP$-violating asymmetry
\begin{equation}
{\cal A}_{K\pi}(t) \; \equiv \; \frac{R(D^0_{\rm phys}(t)\rightarrow K^-\pi^+) ~ - ~
R(\bar{D}^0_{\rm phys}(t)\rightarrow K^+\pi^-)}
{R(D^0_{\rm phys}(t)\rightarrow K^-\pi^+) ~ + ~ R(\bar{D}^0_{\rm phys}(t)\rightarrow K^+\pi^-)} \; .
\end{equation}
Explicitly, we get
\begin{eqnarray}
{\cal A}_{K\pi}(t) & \approx & -\lambda^2 h_{K\pi} \left [ \hat{\Delta}_D \cos(2\phi)
\left (y^{~}_D \cos\delta_{K\pi} + x^{~}_D \sin\delta_{K\pi}\right ) \right . \nonumber \\
&  & \left . + \sin(2\phi) \left ( y^{~}_D \sin\delta^{~}_{K\pi} - 
x^{~}_D \cos\delta^{~}_{K\pi}\right ) \right ] \Gamma t \; ,
\end{eqnarray}
where the observable $\hat{\Delta}_D$ has been defined before (see eq. (4.6)). 
One can see that ${\cal A}_{K\pi}(t)$ are composed of two sources of
$CP$-violating effects, that in $D^0-\bar{D}^0$ mixing (proportional to $\hat{\Delta}_D$)
and that from the interplay of decay and mixing (proportional to $\sin(2\phi))$.
The magnitude of ${\cal A}_{K\pi}(t)$ is constrained by both the DCSD amplitude
$\lambda^2 h_{K\pi}$ and the $D^0-\bar{D}^0$ mixing parameters $x^{~}_D$ and $y^{~}_D$.
Since $\hat{\Delta}_D\leq 10^{-2}$ is expected, ${\cal A}_{K\pi}(t)$ can reach the percent
level only when the $\sin(2\phi)$ term is significantly enhanced by new physics.
In new physics scenarios with $y^{~}_D\ll x^{~}_D$ \cite{Blaylock-Seiden-Nir,Browder-Pakvasa}, 
we get ${\cal A}_{K\pi}(t)\approx \lambda^2 h_{K\pi} x^{~}_D \sin(2\phi)
\cos\delta_{K\pi} \Gamma t$ as a safe approximation. Taking $h_{K\pi}\approx 1.8$
(implied by $|\rho^{~}_{K^-\pi^+}|^2\approx 7.7\times 10^{-3}$ \cite{CLEO}), 
$x^{~}_D< 0.086$ and $\delta_{K\pi}\geq 5^0$, one finds the restriction 
${\cal A}_{K\pi}(t)<7.5\times 10^{-3}\Gamma t$.

\vspace{0.3cm}

(2) The combined decay rate
\begin{equation}
R_{K\pi}(t) \; \equiv \; R(D^0_{\rm phys}(t)\rightarrow K^+\pi^-) ~ + ~
R(\bar{D}^0_{\rm phys}(t)\rightarrow K^-\pi^+) \; .
\end{equation}
By use of eq. (5.5), we obtain
\begin{eqnarray}
R_{K\pi}(t) & \propto & |A_{K^-\pi^+}|^2 \exp(-\Gamma t) \left \{ 
2\lambda^4 h^2_{K\pi} + r^{~}_D \Gamma^2 t^2 \right . \nonumber \\
&  & + 2\lambda^2 h_{K\pi} \left [ \cos(2\phi) \left (y^{~}_D \cos\delta_{K\pi}
- x^{~}_D \sin\delta_{K\pi} \right ) \right . \nonumber \\
&  & \left . \left . - \hat{\Delta}_D \sin(2\phi) \left ( y^{~}_D \sin\delta_{K\pi}
+x^{~}_D \cos\delta_{K\pi} \right ) \right ] \Gamma t \right \} \; ,
\end{eqnarray}
where $r^{~}_D=(r+\bar{r})/2$ defined in eq. (3.3) has been used. The three terms of
$R_{K\pi}(t)$, which have different time distributions, come respectively from DCSD,
$D^0-\bar{D}^0$ mixing, and the interplay of these two effects. Thus the detection of
$R_{K\pi}(t)$ can determine the $D^0-\bar{D}^0$ mixing rate $r^{~}_D$ and distinguish it from
the DCSD contribution. If $|\phi|$ is not large (e.g., in the standard model),
the interference term will be dominated by $\cos(2\phi) (y^{~}_D \cos\delta_{K\pi}
- x^{~}_D \sin\delta_{K\pi})$ due to the smallness of $\hat{\Delta}_D$. In this case, the
information about $y^{~}_D$ might be obtainable if the contribution of $x^{~}_D$ to the
interference term is suppressed by small $\delta_{K\pi}$. To justify the possible magnitude 
of $\phi$, however, one has to combine the measurements of ${\cal A}_{K\pi}(t)$
and $R_{K\pi}(t)$.

\vspace{0.3cm}

Now we take a brief look at the time-independent decay rates of $D^0_{\rm phys}/\bar{D}^0_{\rm phys}
\rightarrow K^{\pm}\pi^{\mp}$. With the help of eqs. (2.24) and (2.25), we obtain
\begin{eqnarray}
{\cal A}_{K\pi} & \equiv & \frac{R(D^0_{\rm phys}\rightarrow K^-\pi^+) ~ - ~
R(\bar{D}^0_{\rm phys}\rightarrow K^+\pi^-)}
{R(D^0_{\rm phys}\rightarrow K^-\pi^+) ~ + ~ R(\bar{D}^0_{\rm phys}\rightarrow K^+\pi^-)} 
\nonumber \\
& \approx & -\lambda^2 h_{K\pi} \left [ \hat{\Delta}_D \cos(2\phi)
\left (y^{~}_D \cos\delta_{K\pi} + x^{~}_D \sin\delta_{K\pi}\right ) \right . \nonumber \\
&  & \left . + \sin(2\phi) \left ( y^{~}_D \sin\delta^{~}_{K\pi} - 
x^{~}_D \cos\delta^{~}_{K\pi}\right ) \right ] \; ,
\end{eqnarray}
i.e., ${\cal A}_{K\pi}$ is approximately equal to the value of ${\cal A}_{K\pi}(t)$
at $t=1/\Gamma$. Similarly, one can calculate another $CP$-violating asymmetry:
\begin{eqnarray}
\bar{\cal A}_{K\pi} & \equiv & \frac{R(D^0_{\rm phys}\rightarrow K^+\pi^-) ~ - ~
R(\bar{D}^0_{\rm phys}\rightarrow K^-\pi^+)}
{R(D^0_{\rm phys}\rightarrow K^+\pi^-) ~ + ~ R(\bar{D}^0_{\rm phys}\rightarrow K^-\pi^+)} \nonumber \\
\nonumber \\
& \approx & \lambda^2 h_{K\pi} \left [ -\hat{\Delta}_D \cos(2\phi) \left (y^{~}_D \cos\delta_{K\pi}
- x^{~}_D \sin\delta_{K\pi} \right ) \right . \nonumber \\
&  & \left . + \sin(2\phi) \left ( y^{~}_D \sin\delta_{K\pi}
+ x^{~}_D \cos\delta_{K\pi} \right ) \right ] \left \{ \lambda^4 h^2_{K\pi} \right . \nonumber \\
&  & + r^{~}_D  + \lambda^2 h_{K\pi} \left [ \cos(2\phi) 
\left (y^{~}_D \cos\delta_{K\pi} - x^{~}_D \sin\delta_{K\pi} \right ) \right . \nonumber \\
&  & \left . \left . - \hat{\Delta}_D \sin(2\phi) \left ( y^{~}_D \sin\delta_{K\pi}
+x^{~}_D \cos\delta_{K\pi} \right ) \right ] \right \}^{-1} \; .
\end{eqnarray}
Taking $\hat{\Delta}_D\approx 0$ and $\sin(2\phi)\approx \pm 1$ for example,
we find that $\bar{\cal A}_{K\pi}$ may be significant:
\begin{equation}
\bar{\cal A}_{K\pi} \; \approx \; \pm \frac{ \lambda^2 h_{K\pi} \left ( y^{~}_D \sin\delta_{K\pi}
+ x^{~}_D \cos\delta_{K\pi} \right )}
{\lambda^4 h^2_{K\pi} + r^{~}_D} \; . 
\end{equation}
Due to the suppressed rates of $D^0_{\rm phys}\rightarrow K^+\pi^-$ and
$\bar{D}^0_{\rm phys}\rightarrow K^-\pi^+$, however, the measurement of
$\bar{\cal A}_{K\pi}$ will be a stiff experimental challenge.

\begin{center}
{\large\bf B. ~ Coherent $D$ decays to $K^{\pm}\pi^{\mp}$}
\end{center}

At the $\psi(3.77)$ and $\psi(4.14)$ resonances, the $K^{\pm}\pi^{\mp}$ events may
come from the coherent decays of $(D^0_{\rm phys}\bar{D}^0_{\rm phys})_C$ pairs. The flavor of
one $D$ meson decaying to $K^{\pm}\pi^{\mp}$ can be tagged by detecting the other $D$
decaying to the semileptonic states $l^{\pm}X^{\mp}$. The overall rates for 
such joint decay events, up to $O(x^2_D)$, $O(y^2_D)$ or $O(\lambda^4)$ for every
distinctive term, are obtainable from eqs. (2.26) and (2.27) as follows 
\footnote{The formulas with the assumption of $y^{~}_D \ll x^{~}_D$ and
$|q/p|=1$ have been given in ref. \cite{Xing2}.}: 
\begin{eqnarray}
R(l^-, K^-\pi^+)_- & \propto & |A_l|^2 |A_{K^-\pi^+}|^2 \left ( 2 - x^2_D + y^2_D
\right ) \; , \nonumber \\ 
R(l^+, K^+\pi^-)_- & \propto & |A_l|^2 |A_{K^-\pi^+}|^2 \left ( 2 - x^2_D + y^2_D
\right ) \; , \nonumber \\
R(l^-, K^+\pi^-)_- & \propto & |A_l|^2 |A_{K^-\pi^+}|^2 \left [ 2 \lambda^4 h^2_{K\pi}
+ \left (x^2_D + y^2_D\right ) \left | \frac{q}{p} \right |^2 \right ] \; , \nonumber \\
R(l^+, K^-\pi^+)_- & \propto & |A_l|^2 |A_{K^-\pi^+}|^2 \left [ 2 \lambda^4 h^2_{K\pi}
+ \left (x^2_D + y^2_D\right ) \left | \frac{p}{q} \right |^2 \right ] \; ;
\end{eqnarray}
and
\begin{eqnarray}
R(l^-, K^-\pi^+)_+ & \propto & |A_l|^2 |A_{K^-\pi^+}|^2 \left \{ 2 - 3 \left (
x^2_D - y^2_D \right ) \right . \nonumber \\
&  & \left . + ~ 4 \lambda^2 h_{K\pi} \left | \frac{q}{p} \right | 
\left [ y^{~}_D \cos (\delta_{K\pi} +2\phi) + x^{~}_D \sin (\delta_{K\pi}
+ 2\phi) \right ] \right \} \; , \nonumber \\
R(l^+, K^+\pi^-)_+ & \propto & |A_l|^2 |A_{K^-\pi^+}|^2 \left \{ 2 - 3 \left (
x^2_D - y^2_D \right ) \right . \nonumber \\
&  & \left . + ~ 4 \lambda^2 h_{K\pi} \left | \frac{p}{q} \right | 
\left [ y^{~}_D \cos (\delta_{K\pi} -2\phi) + x^{~}_D \sin (\delta_{K\pi}
- 2\phi) \right ] \right \} \; , \nonumber \\
R(l^-, K^+\pi^-)_+ & \propto & |A_l|^2 |A_{K^-\pi^+}|^2 \left \{ 2 \lambda^4 h^2_{K\pi}
+ 3 \left (x^2_D + y^2_D \right ) \left | \frac{q}{p} \right |^2 \right . \nonumber \\
&  & \left . + ~ 4 \lambda^2 h_{K\pi} \left | \frac{q}{p} \right |
\left [ y^{~}_D \cos (\delta_{K\pi} -2\phi) - x^{~}_D \sin (\delta_{K\pi}
- 2\phi) \right ] \right \} \; , \nonumber \\
R(l^+, K^-\pi^+)_+ & \propto & |A_l|^2 |A_{K^-\pi^+}|^2 \left \{ 2 \lambda^4 h^2_{K\pi} 
+ 3 \left (x^2_D + y^2_D \right ) \left | \frac{p}{q} \right |^2 \right . \nonumber \\
&  & \left . + ~ 4 \lambda^2 h_{K\pi} \left | \frac{p}{q} \right |
\left [ y^{~}_D \cos (\delta_{K\pi} +2\phi) - x^{~}_D \sin (\delta_{K\pi}
+ 2\phi) \right ] \right \} \; .
\end{eqnarray}
Some discussions about these results are in order.

\vspace{0.3cm}

(1) To an excellent degree of accuracy, we have
\begin{equation}
R(l^-, K^-\pi^+)_- \; \approx \; R(l^+, K^+\pi^-)_- \; .
\end{equation}
The joint decay rates $R(l^{\mp}, K^{\pm}\pi^{\mp})_-$ can be normalized by
$R(l^{\mp}, K^{\mp}\pi^{\pm})_-$, and the resultant rate difference or sum reads
\begin{eqnarray}
S^{(-)}_- & \equiv & \frac{R(l^-, K^+\pi^-)_-}{R(l^-, K^-\pi^+)_-}
~ - ~ \frac{R(l^+, K^-\pi^+)_-}{R(l^+, K^+\pi^-)_-} \; \approx \;
- 2 r^{~}_D \Delta_D \; , \nonumber \\
S^{(+)}_- & \equiv & \frac{R(l^-, K^+\pi^-)_-}{R(l^-, K^-\pi^+)_-}
~ + ~ \frac{R(l^+, K^-\pi^+)_-}{R(l^+, K^+\pi^-)_-} \; \approx \;
2 \lambda^4 h^2_{K\pi} + 2 r^{~}_D \; ,
\end{eqnarray}
where $r^{~}_D$ and $\Delta_D$ have been given in eqs. (3.3) and (3.6), respectively. Observation of the
$CP$-violating asymmetry $S^{(-)}_-$ may be practically impossible due to the smallness
of $\Delta_D$ and $r^{~}_D$. However, $S^{(+)}_-$ is expected to be measurable
at a $\tau$-charm factory running on the $\psi(3.77)$ resonance. As we shall show 
in the next subsection, $r^{~}_D$ can be extracted from the joint decay rates
$R(K^+\pi^-, K^+\pi^-)_-$ and $R(K^-\pi^+, K^+\pi^-)_-$, thus a 
comparison of this measurement with that for $S^{(+)}_-$ will separately 
determine the magnitudes of $D^0-\bar{D}^0$ mixing and DCSD. This idea is interesting
on the point that the relevant measurements are time-independent and the involved 
decay modes are only $D^0/\bar{D}^0\rightarrow K^{\pm}\pi^{\mp}$.

\vspace{0.3cm}

(2) It is easy to obtain the rate asymmetry
\begin{equation}
\frac{R(l^-, K^-\pi^+)_+ - R(l^+, K^+\pi^-)_+}{R(l^-, K^-\pi^+)_+ +
R(l^+, K^+\pi^-)_+} \; \approx \; 2 {\cal A}_{K\pi} \; ,
\end{equation}
where ${\cal A}_{K\pi}$ has been given in eq. (5.10). Normalizing the joint decay 
rates $R(l^{\mp}, K^{\pm}\pi^{\mp})_+$ by $R(l^{\mp}, K^{\mp}\pi^{\pm})_+$, we get
\begin{eqnarray}
S^{(-)}_+ & \equiv & \frac{R(l^-, K^+\pi^-)_+}{R(l^-, K^-\pi^+)_+}
~ - ~ \frac{R(l^+, K^-\pi^+)_+}{R(l^+, K^+\pi^-)_+} \nonumber \\
& \approx & - 6 r^{~}_D \Delta_D - 4 \lambda^2 h_{K\pi} \left [ \hat{\Delta}_D
\cos(2\phi) \left ( y^{~}_D \cos\delta_{K\pi} - x^{~}_D \sin\delta_{K\pi}
\right ) \right . \nonumber \\
&  & \left . - \sin(2\phi) \left ( y^{~}_D \sin\delta_{K\pi} +
x^{~}_D \cos\delta_{K\pi} \right ) \right ] \; , \nonumber \\ \nonumber \\
S^{(+)}_+ & \equiv & \frac{R(l^-, K^+\pi^-)_+}{R(l^-, K^-\pi^+)_+}
~ + ~ \frac{R(l^+, K^-\pi^+)_+}{R(l^+, K^+\pi^-)_+} \nonumber \\
& \approx & 2 \lambda^4 h^2_{K\pi} + 6 r^{~}_D 
+ 4 \lambda^2 h_{K\pi} \left [ \cos(2\phi) \left ( y^{~}_D \cos\delta_{K\pi}
- x^{~}_D \sin\delta_{K\pi} \right ) \right . \nonumber \\
&  & \left . - \hat{\Delta}_D \sin(2\phi) \left ( y^{~}_D \sin\delta_{K\pi}
+ x^{~}_D \cos\delta_{K\pi} \right ) \right ] \; .
\end{eqnarray}
From eqs. (5.11), (5.16) and (5.18), one can see the following relation among $S^{(\pm)}_{\pm}$
and $\bar{\cal A}_{K\pi}$:
\begin{equation}
\bar{\cal A}_{K\pi} \; \approx \; \frac{S^{(-)}_+  -  3 S^{(-)}_-}
{S^{(+)}_+  -  4 r^{~}_D  +  S^{(+)}_-} \; .
\end{equation}
This result could be tested if the data on all six measurables were available.

\vspace{0.3cm}

(3) To give one a feeling of ballpark numbers to be expected, we roughly estimate
the magnitudes of the above-mentioned observables by assuming $\Delta_D = \hat{\Delta}_D =0$
and $y^{~}_D \ll x^{~}_D$.
Taking the semileptonic decay mode serving for flavor tagging to be $D^0\rightarrow
K^-e^+\nu_e$, we have its branching ratio $B(D^0\rightarrow K^-e^+\nu_e)\approx
3.8\%$ \cite{PDG}. In addition, the current data give $B(D^0\rightarrow K^-\pi^+)
\approx 4.01\%$ \cite{PDG}. Then $R(l^+, K^+\pi^-)_{\pm}$ and $R(l^-, K^-\pi^+)_{\pm}$
are at the level $10^{-3}$ or so, while $R(l^-, K^+\pi^-)_{\pm}$ and
$R(l^+, K^-\pi^+)_{\pm}$ may be of the order $10^{-5}$ if we input $x^{~}_D
\sim 0.06$. Within the experimental capabilities of a $\tau$-charm factory, it is
possible to measure the latter four decay rates to an acceptable degree of accuracy
with about $10^7$ $D^0\bar{D}^0$ events \cite{Xing2}. Furthermore, the upper bounds of 
the $CP$ asymmetries ${\cal A}_{K\pi}$ and $S^{(-)}_+$ can be obtained by use of
the experimental results $x^{~}_D < 0.086$ and $|\rho^{~}_{K^-\pi^+}|^2
\approx 0.77\%$ \cite{PDG,CLEO}. Taking $\cos\delta_{K\pi}=1$ and $\sin(2\phi)
=\pm 1$, we get $|{\cal A}_{K\pi}| < 0.008$ and $|S^{(-)}_+| < 0.03$. 
In the assumption of perfect detectors or $100\%$ tagging efficiencies, one needs
about $10^8$ $D^0\bar{D}^0$ events to uncover $|S^{(-)}_+|\sim 0.01$ at the level
of three standard deviations or to measure $2|{\cal A}_{K\pi}| \sim 0.005$
in eq. (5.17) at the level of one standard deviation. Accumulation of so many events
is of course a serious challenge to all types of experimental facilities for charm physics, but 
it should be achievable in the second-round experiments of a $\tau$-charm factory.

\begin{center}
{\large\bf C. ~ Ratios of $R(K^{\pm}\pi^{\mp}, K^{\pm}\pi^{\mp})_C$ to
$R(K^{\pm}\pi^{\mp}, K^{\mp}\pi^{\pm})_C$}
\end{center}

It has been pointed out that the coherent decays $(D^0_{\rm phys}\bar{D}^0_{\rm phys})_C
\rightarrow (K^{\pm}\pi^{\mp}) (K^{\pm}\pi^{\mp})$ can be used to search for 
$D^0-\bar{D}^0$ mixing and to separate it from the DCSD effect \cite{Yamamoto}.
The relevant measurables are
\begin{equation}
r^{+-}_C \; \equiv \; \frac{R(K^+\pi^-, K^+\pi^-)_C}{R(K^-\pi^+, K^+\pi^-)_C} \; ,
~~~~~~~~
r^{-+}_C \; \equiv \; \frac{R(K^-\pi^+, K^-\pi^+)_C}{R(K^-\pi^+, K^+\pi^-)_C} \; .
\end{equation}
Since in previous calculations the effects of $CP$ violation or nonvanishing 
$\delta_{K\pi}$ on $r^{\pm\mp}_C$ were neglected, it is worth having a recalculation
for these observables without special approximations.

\vspace{0.3cm}

By use of eqs. (2.26), (2.27) and (5.3), we obtain
\begin{eqnarray}
R(K^-\pi^+, K^+\pi^-)_- & \propto & |A_{K^-\pi^+}|^4 \left [ 2 - x^2_D + y^2_D 
- 4 \lambda^4 h_{K\pi}^2 \cos(2\delta_{K\pi}) \right ] \; , \nonumber \\
R(K^+\pi^-, K^-\pi^+)_- & \propto & |A_{K^-\pi^+}|^4 \left [ 2 - x^2_D + y^2_D
- 4 \lambda^4 h_{K\pi}^2 \cos(2\delta_{K\pi}) \right ] \; , \nonumber \\
R(K^-\pi^+, K^-\pi^+)_- & \propto & |A_{K^-\pi^+}|^4 \left (x^2_D + y^2_D \right )
\left | \frac{p}{q} \right |^2 \; , \nonumber \\
R(K^+\pi^-, K^+\pi^-)_- & \propto & |A_{K^-\pi^+}|^4 \left (x^2_D + y^2_D \right )
\left | \frac{q}{p} \right |^2 \; ; 
\end{eqnarray}
and
\begin{eqnarray}
R(K^-\pi^+, K^+\pi^-)_+ & \propto & |A_{K^-\pi^+}|^4 \left \{ 2 - 3 x^2_D + 3 y^2_D 
+ 4 \lambda^4 h_{K\pi}^2 \cos(2\delta_{K\pi}) \right . \nonumber \\
&  & + 4 \lambda^2 h_{K\pi} \left | \frac{q}{p} \right | \left [ y^{~}_D \cos \left (\delta_{K\pi}
+ 2\phi \right ) + x^{~}_D \sin \left (\delta_{K\pi} + 2\phi \right ) 
\right ] \; \nonumber \\
&  & \left . + 4 \lambda^2 h_{K\pi} \left | \frac{p}{q} \right | \left [ y^{~}_D \cos \left (\delta_{K\pi}
- 2\phi \right ) + x^{~}_D \sin \left (\delta_{K\pi} - 2\phi \right ) 
\right ] \right \} \; , \nonumber \\
R(K^+\pi^-, K^-\pi^+)_+ & \propto & |A_{K^-\pi^+}|^4 \left \{ 2 - 3 x^2_D + 3 y^2_D
+ 4 \lambda^4 h_{K\pi}^2 \cos(2\delta_{K\pi}) \right . \nonumber \\
&  & + 4 \lambda^2 h_{K\pi} \left | \frac{q}{p} \right | \left [ y^{~}_D \cos \left (\delta_{K\pi}
+ 2\phi \right ) + x^{~}_D \sin \left (\delta_{K\pi} + 2\phi \right )
\right ] \; \nonumber \\
&  & \left . + 4 \lambda^2 h_{K\pi} \left | \frac{p}{q} \right | \left [ y^{~}_D \cos \left (\delta_{K\pi}
- 2\phi \right ) + x^{~}_D \sin \left (\delta_{K\pi} - 2\phi \right )
\right ] \right \} \; , \nonumber \\
R(K^-\pi^+, K^-\pi^+)_+ & \propto & |A_{K^-\pi^+}|^4 \left \{ 3 \left (x^2_D + y^2_D \right )
\left | \frac{p}{q} \right |^2 + 8 \lambda^4 h_{K\pi}^2 \right . \nonumber \\
&  & \left . + ~ 8 \lambda^2 h_{K\pi} \left | \frac{p}{q} \right | \left [ y^{~}_D \cos \left (\delta_{K\pi} +
2 \phi \right ) - x^{~}_D \sin \left (\delta_{K\pi} + 2\phi \right )
\right ] \right \} \; , \nonumber \\
R(K^+\pi^-, K^+\pi^-)_+ & \propto & |A_{K^-\pi^+}|^4 \left \{ 3 \left (x^2_D + y^2_D \right )
\left | \frac{q}{p} \right |^2 + 8 \lambda^4 h_{K\pi}^2 \right . \nonumber \\
&  & \left . + ~ 8 \lambda^2 h_{K\pi} \left | \frac{q}{p} \right | \left [ y^{~}_D \cos \left (\delta_{K\pi} -
2 \phi \right ) - x^{~}_D \sin \left (\delta_{K\pi} - 2\phi \right )
\right ] \right \} \; 
\end{eqnarray}
up to $O(x^2_D)$, $O(y^2_D)$ or $O(\lambda^4)$. Clearly $R(K^-\pi^+, K^+\pi^-)_C \approx R(K^+\pi^-, K^-\pi^+)_C$ 
holds to an excellent degree of accuracy. As a consequence, the ratios $r^{\pm\mp}_C$ are given by
\begin{equation}
r^{+-}_- \; \approx \; \frac{x^2_D + y^2_D}{2} \left | \frac{q}{p} \right |^2 \; , ~~~~~~~~
r^{-+}_- \; \approx \; \frac{x^2_D + y^2_D}{2} \left | \frac{p}{q} \right |^2 \; ; 
\end{equation}
and
\begin{eqnarray}
r^{+-}_+ & \approx & 3 r^{+-}_- + 4\lambda^4 h^2_{K\pi} + 4 \lambda^2 h_{K\pi}
\left | \frac{q}{p} \right | \nonumber \\
&  & \times \left [ y^{~}_D \cos \left (\delta_{K\pi} - 2\phi \right )
- x^{~}_D \sin \left (\delta_{K\pi} - 2\phi \right ) \right ] \; , \nonumber \\
r^{-+}_+ & \approx & 3 r^{-+}_- + 4\lambda^4 h^2_{K\pi} + 4 \lambda^2 h_{K\pi}
\left | \frac{p}{q} \right | \nonumber \\
&  & \times \left [ y^{~}_D \cos \left (\delta_{K\pi} + 2\phi \right )
- x^{~}_D \sin \left (\delta_{K\pi} + 2\phi \right ) \right ] \; . 
\end{eqnarray}
One can see that $r^{+-}_-$ and $r^{-+}_-$ are approximately equivalent to $r$ and $\bar{r}$ 
obtained in eq. (3.2). The difference between $r^{+-}_-$ and $r^{-+}_-$ measures
$CP$ violation in $D^0-\bar{D}^0$ mixing, and the sum of them amounts approximately
to $r^{~}_D$ given in eq. (3.3). The DCSD effect on $r^{+-}_+$ and $r^{-+}_+$
is significant and non-negligible, but its magnitude can be isolated from the
difference $r^{+-}_+ - 3 r^{+-}_-$ or $r^{-+}_+ - 3 r^{-+}_-$. In addition, we
find
\begin{eqnarray}
r^{-+}_+ ~ - ~ r^{+-}_+ & \approx & 8 \lambda^2 h_{K\pi} \left [ \hat{\Delta}_D
\cos(2\phi) \left (y^{~}_D \cos\delta_{K\pi} - x^{~}_D \sin\delta_{K\pi}
\right ) \right . \nonumber \\
&  & \left . - \sin(2\phi) \left (y^{~}_D \sin\delta_{K\pi} + 
x^{~}_D \cos\delta_{K\pi} \right ) \right ] \; .
\end{eqnarray}
Comparing this result with those derived in eq. (5.18), one gets 
\begin{equation}
r^{-+}_+ ~ - ~ r^{+-}_+ \; \approx \; -2 \left (6 r^{~}_D \Delta_D ~ + ~ 
S^{(-)}_+ \right ) \; .
\end{equation}
Such a $CP$-violating signal might be detectable at a $\tau$-charm factory running on the
$\psi(4.14)$ resonance.

\vspace{0.3cm}

Although the above discussions concentrate only on $D^0/\bar{D}^0\rightarrow 
K^{\pm}\pi^{\mp}$, the similar results can be obtained for some other decay
modes taking place via the same quark diagrams, such as $D^0\rightarrow
K^{\pm}\rho^{\mp}$, $K^{*\pm}\pi^{\mp}$ and their flavor-conjugate processes.
All these channels are expected to have the same weak interactions, but their 
final-state interactions may be different from one another (e.g., $\delta_{K\pi}
\neq \delta_{K\rho}$). If the $SU(3)$ breaking effects in 
$D^0/\bar{D}^0\rightarrow (K^{\pm}, K^{*\pm}) + (\pi^{\mp}, \rho^{\mp}, a^{\mp}_1,
~ {\rm etc})$ are not so significant that all the strong phase shifts lie in the
same quadrant as $\delta_{K\pi}$, then a sum over these modes is possible to
increase the number of decay events in statistics, with few dilution effect
on the signal of $D^0-\bar{D}^0$ mixing and $CP$ violation.

\section{On $CP$-forbidden decays}
\setcounter{equation}{0}

We now consider $CP$-forbidden transitions of the type
\begin{equation}
(D^0_{\rm phys}\bar{D}^0_{\rm phys})_{\pm} \; \rightarrow \; (f_1 f_2)_{\mp} \; ,
\end{equation}
where the $D^0\bar{D}^0$ pair with definite $CP$ parity can be coherently
produced on the $\psi(3.77)$ or $\psi(4.14)$ resonance, $f_1$ and $f_2$ denote
the $CP$ eigenstates with the same or opposite $CP$ parity. It is worth
remarking that for such decay modes the $CP$-violating signals can be established
by detecting the joint decay rates other than the decay-rate asymmetries. In
practice, this implies that neither flavor-tagging for the initial $D$ mesons
nor time-dependent measurements of the whole decay processes are necessary.
The joint decay rate $R(f_1, f_2)_C$ and its analytical approximation have been
presented in eqs. (2.21) and (2.26). For simplicity and illustration, here we
concentrate mainly on the $CP$-forbidden decays $(D^0_{\rm phys}\bar{D}^0_{\rm phys})_-
\rightarrow (f_1 f_2)_+$, such as $(f_1 f_2)_+ = (K^+K^-)(\pi^+\pi^-)$ and
$(K^+K^-)(K^+K^-)$. The case $(D^0_{\rm phys}\bar{D}^0_{\rm phys})_+ \rightarrow
(f_1 f_2)_-$ will be briefly discussed by taking $f_1 = K_S \pi^0$ and $f_2 = K_L \pi^0$
for example.

\vspace{0.3cm}

By use of the quantities ${\cal U}_f$, ${\cal V}_f$ and ${\cal W}_f$ defined in
eq. (4.2), the joint decay rate $R(f_1, f_2)_-$ can be written as
\begin{eqnarray}
R(f_1, f_2)_- & \propto & |A_{f_1}|^2 |A_{f_2}|^2 \left (1+|\lambda_{f_1}|^2 \right )
\left (1+|\lambda_{f_2}|^2\right ) \left | \frac{p}{q} \right |^2 \times \nonumber \\
&  & \left [ \frac{1}{1-y^2_D} \left (1-{\cal W}_{f_1} {\cal W}_{f_2} \right )
- \frac{1}{1+x^2_D} \left ( {\cal U}_{f_1} {\cal U}_{f_2} + {\cal V}_{f_1}
{\cal V}_{f_2} \right ) \right ] \; .
\end{eqnarray}
Here we assume $f_1$ and $f_2$ to be two $CP$ eigenstates with the same $CP$
parity. $CP$ conservation requires ${\cal V}_{f_1}={\cal V}_{f_2}=0$,
${\cal U}_{f_1}={\cal U}_{f_2}=0$ and ${\cal W}_{f_1}={\cal W}_{f_2}=\pm 1$,
then we get $R(f_1, f_2)_- = 0$. Thus nonvanishing $R(f_1, f_2)_-$ is a clean
signal of $CP$ violation. In the special case $f_1 = f_2 \equiv f$, one finds
\begin{equation}
R(f, f)_- \; \propto \; |A_f|^4 \left (1+|\lambda_f|^2 \right )^2
\left | \frac{p}{q} \right |^2 \left ( \frac{1}{1-y^2_D} - \frac{1}{1+x^2_D} \right )
\left ( {\cal U}^2_f + {\cal V}^2_f \right ) \; .
\end{equation}
This result can be straightforwardly obtained from eq. (6.2) with the help
of eq. (4.3). As discussed before, ${\cal U}_f$ is composed of the $CP$ asymmetry in
$D^0-\bar{D}^0$ mixing and that in the direct transition amplitudes of $D$
decays, while ${\cal V}_f$ signifies the $CP$ asymmetry induced by the
interplay of decay and $D^0-\bar{D}^0$ mixing. Due to the smallness of
${\cal U}_f$, ${\cal V}_f$, $x^{~}_D$ and $y^{~}_D$, we believe that
$R(f, f)_-$ must be significantly suppressed.

\vspace{0.3cm}

In comparison with $R(f, f)_-$, the joint decay rate $R(f, f)_+$ is not
$CP$-forbidden:
\begin{eqnarray}
R(f, f)_+ & \propto & |A_f|^4 \left (1+|\lambda_f|^2 \right )^2 \left | \frac{p}{q} \right |^2
\left [ \frac{1+y^2_D}{\left (1-y^2_D\right )^2} \left (1+{\cal W}_f \right )^2
\right . \nonumber \\
&  & \left . - \frac{4y^{~}_D}{\left (1-y^2_D\right )^2} {\cal W}_f 
- \frac{1-x^2_D}{\left (1+x^2_D\right )^2} \left ({\cal U}^2_f -
{\cal V}^2_f \right ) - \frac{4x^{~}_D}{\left (1+x^2_D\right )^2}
{\cal U}_f {\cal V}_f \right ] \; .
\end{eqnarray}
Approximately, we obtain
\begin{equation}
\frac{R(f, f)_-}{R(f, f)_+} \; \approx \; \frac{\left (x^2_D + y^2_D \right )
\left ( {\cal U}_f^2 + {\cal V}_f^2 \right )}{1+ {\cal W}^2_f -
4 y^{~}_D {\cal W}_f} \; .
\end{equation}
This relation can in principle be tested for $f = K^+K^-$ etc at the 
$\psi(4.14)$ resonance in the second-round experiments of a $\tau$-charm
factory, if the rate of $D^0-\bar{D}^0$ mixing is at the detectable level.

\vspace{0.3cm}

In the neglect of $CP$ violation in $K^0-\bar{K}^0$ mixing, the states 
$K_S \pi^0$ and $K_L \pi^0$ are two $CP$ eigenstates with the opposite
$CP$ parity. Thus the process $(D^0_{\rm phys}\bar{D}^0_{\rm phys})_+
\rightarrow (K_S \pi^0) (K_L \pi^0)$ should be $CP$-forbidden.
As a good approximation, we have $|A_{K_L \pi^0}| \approx |A_{K_S \pi^0}|$
and $\rho^{~}_{K_L \pi^0} \approx - \rho^{~}_{K_S \pi^0}$ (see eq. (4.22)).
Then the joint decay rate with $C=+$ turns out to be
\begin{eqnarray}
R(K_L \pi^0, K_S \pi^0)_+ & \propto & |A_{K_S \pi^0}|^4 \left (1+|\lambda_{K_S \pi^0}|^2
\right )^2 \left | \frac{p}{q} \right |^2 \times \nonumber \\
&  & \left [ \frac{1+y^2_D}{\left (1-y^2_D\right )^2} - \frac{1-x^2_D}
{\left (1+x^2_D\right )^2} \right ] \left ({\cal U}_{K_S \pi^0}^2
+ {\cal V}_{K_S \pi^0}^2 \right ) \; .
\end{eqnarray}
Using the approximate results in eq. (4.24) and taking $|\epsilon|\approx 0$,
we obtain a simpler expression for the equation above:
\begin{equation}
R(K_L \pi^0, K_S \pi^0)_+ \; \propto \; 6 |A_{K_S \pi^0}|^4 ( 1 + w)
\left (x^2_D + y^2_D \right ) \left [ \hat{\Delta}^2_D + \sin^2 (2\phi) \right ] \; ,
\end{equation}
where $w$ has been defined in eq. (3.3).
In contrast, it is easy to check from eq. (6.3) that
\begin{equation}
R(K_L \pi^0, K_L \pi^0)_- \; \approx \; R(K_S \pi^0, K_S \pi^0)_- \; \approx \;
\frac{1}{3} R(K_L \pi^0, K_S \pi^0)_+ \; .
\end{equation}
Note that $CP$ violation in $D^0-\bar{D}^0$ mixing (i.e., $\hat{\Delta}_D$) might
be negligibly small, thus the dominant signal of $CP$ violation in $R(K_S \pi^0,
K_S \pi^0)_-$ or $R(K_L \pi^0, K_S \pi^0)_+$ could come from the mixing phase
$\phi$ enhanced by new physics. In this sense, it is worthwhile to
experimentally search for the above-mentioned $CP$-forbidden transitions.

\section{Summary}
\setcounter{equation}{0}

To meet various delicate experiments in the near future at fixed target machines,
$B$-meson factories and $\tau$-charm factories, we have made a further study of 
the phenomenology of $D^0-\bar{D}^0$ mixing and $CP$ violation in neutral $D$-meson
decays. The generic formulas for the time-dependent and time-integrated decay rates
of both coherent and incoherent $D^0\bar{D}^0$ events were derived, and their 
approximate expressions up to the accuracy of $O(x^2_D)$ and $O(y^2_D)$ were
presented. A variety of $D^0-\bar{D}^0$ mixing and $CP$-violating signals were
analyzed in detail for neutral $D$ decays to the semileptonic states, the
nonleptonic $CP$ eigenstates, the nonleptonic non-$CP$ eigenstates and the 
$CP$-forbidden states. 

\vspace{0.3cm}

In particular, we have shown that it is possible to separately determine the 
magnitudes of $x^{~}_D$ and $y^{~}_D$ through precise measurements of the dilepton
events of coherent $D^0\bar{D}^0$ decays on the $\psi(4.14)$ resonance at a $\tau$-charm 
factory. We gave a detailed analysis of $D^0-\bar{D}^0$ mixing signals
and DCSD effects in the time-dependent and time-independent decays $D^0/\bar{D}^0\rightarrow
K^{\pm}\pi^{\mp}$. It is found that some constraints on $x^{~}_D$ and $y^{~}_D$
can be achieved in both fixed target and $\tau$-charm factory experiments,
and the mixing and DCSD effects are distinguishable from each other. 
Taking $CP$ violation and final-state interactions into account, we recalculated 
the joint decay rates of coherent $D^0\bar{D}^0$ pairs to $(K^{\pm}\pi^{\mp})(K^{\pm}\pi^{\mp})$,
which are useful for the time-independent determination of $r^{~}_D$ and
DCSD amplitudes. A special attention has been paid to the $D^0-\bar{D}^0$
mixing signals in the decay modes $D^0/\bar{D}^0\rightarrow K_{S,L} + \pi^0$ etc.
We pointed out that a model-independent restriction on $x^{~}_D$ and $y^{~}_D$ 
should be obtainable from the time distributions of such decay modes.

\vspace{0.3cm}

$CP$ violation in $D^0-\bar{D}^0$ mixing can be well constrained in the semileptonic
decays of coherent or incoherent $D^0\bar{D}^0$ events. In addition to this source
of $CP$ asymmetry, we have shown that both the direct $CP$ asymmetry in the transition amplitudes
of $D$ decays and the indirect $CP$ asymmetry arising from the interplay of decay and
$D^0-\bar{D}^0$ mixing can also manifest themselves in neutral $D$ decays to hadronic $CP$
eigenstates. These different $CP$-violating signals usually have different
time distributions in the decay rates, thus they are possible to be distinguished from
one another. In particular, direct $CP$ violation can be cleanly probed in the coherent
$(D^0\bar{D}^0)_-$ decays to a $CP$ eigenstate plus a semileptonic state
on the $\psi(3.77)$ or $\psi(4.14)$ resonance. For the decay modes with $K^0-\bar{K}^0$
mixing in the final states, however, the $CP$ asymmetry induced by the mixing
parameter $\epsilon$ may be
non-negligible and even dominant over the direct $CP$-violating signal from the 
charm quark transitions. Taking $D\rightarrow K\bar{K}$ for example, we illustrated
the significant effects of final-state interactions on $CP$ violation. 
Different from those neutral $D$ decays to $CP$ eigenstates, $D^0/\bar{D}^0
\rightarrow K^{\pm}\pi^{\mp}$ are expected to have no direct $CP$ asymmetries.
Although the indirect $CP$-violating effects exist in such processes, they are suppressed 
to some extent by the DCSD amplitudes. We also discussed the $CP$-forbidden
transitions on the $\psi(3.77)$ and $\psi(4.14)$ resonances. A search for the
$CP$-forbidden modes like $(D^0\bar{D}^0)_-\rightarrow (K^+K^-)(\pi^+\pi^-)$
and $(D^0\bar{D}^0)_+\rightarrow (K_S\pi^0)(K_L\pi^0)$ is worthwhile
in the future experiments of charm physics.

\vspace{0.3cm}

Throughout our calculations $CPT$ symmetry in the $D^0-\bar{D}^0$ mixing
matrix has been assumed. Also the $\Delta Q = \Delta C$ rule was assumed to
hold in most cases, but the effects of $\Delta Q = - \Delta C$ transitions
on $D^0-\bar{D}^0$ mixing and $CP$ violation were briefly discussed in
section 3-B. Due to the smallness of $x^{~}_D$ and $y^{~}_D$, it will be
very difficult to accurately test the $\Delta Q = \Delta C$ rule and 
$CPT$ invariance in the $D^0-\bar{D}^0$ system. Recently Colladay and
Kosteleck$\rm\acute{y}$ have studied a few possibilities to examine 
$CPT$ symmetry in neutral $D$ decays on the basis of the future fixed target
and $\tau$-charm factory experiments \cite{Colladay-Kostelecky}.
Considering this work and some other works on tests of $CPT$ symmetry in
the $B^0-\bar{B}^0$ system \cite{Kobayashi-Sanda,XingCPT}, we want to remark that one of the most sensitive
signals for $CPT$ violation or $\Delta Q = - \Delta C$ transitions should be the
nonvanishing asymmetry $\bar{\Delta}_D$ defined in eq. (3.5). 
However, one should keep in mind that $\bar{\Delta}_D \neq 0$ might also come from
the phase shifts of final-state electromagnetic interactions or the $CP$-violating 
contributions of non-standard electroweak models to the tree-level $W$-mediated semileptonic
$D$ decays. 
Another possible way to test $CPT$ invariance in $D^0-\bar{D}^0$ mixing, which in principle works,
is to measure the time distributions of opposite-sign 
dilepton events at an asymmetric $\tau$-charm factory (see Appendix B).

\vspace{0.3cm}

Of course much more theoretical effort should be made to 
give reliable numerical predicitions for the magnitudes of various $D^0-\bar{D}^0$
mixing and $CP$-violating phenomena. 

\vspace{0.3cm}

\begin{flushleft}
{\Large\bf Acknowledgements}
\end{flushleft}

I would like to thank A.I. Sanda for his warm hospitality and
the Japan Society for the Promotion of Science for its financial support.
In particular, I am grateful to A.I. Sanda for his enlightening 
comments and constructive suggestions on part of this work.

\vspace{0.5cm}

\begin{flushleft}
\Large\bf Appendix A
\end{flushleft}

This appendix is devoted to giving some generic formulas for the time-dependent
$D$ decays at an assumed asymmetric $\tau$-charm factory.
The asymmetric $e^{+}e^{-}$ collisions just above the production threshold of 
$(D^0_{\rm phys}\bar{D}^0_{\rm phys})_C$ pairs
will offer the possibility to measure the decay-time difference
$t_{-}=(t_{2}-t_{1})$ between $D^0_{\rm phys}\rightarrow f_1$ and $\bar{D}^0_{\rm phys}\rightarrow f_2$. 
Usually it is difficult to measure the
$t_{+}=(t_{2}+t_{1})$ distribution in either linacs or storage rings,
unless the bunch lengths are much shorter than the decay
lengths \cite{Feldman-Berkelman}. Here we calculate the $t_{-}$ distributions
of joint decay rates starting from the master formula in eq. (2.19). 
For simplicity, we use $t$ to denote $t_{-}$ in the following.
Integrating $R(f_{1},t_{1};f_{2},t_{2})_{C}$ over $t_{+}$, we obtain
the decay rates (for $C=\pm $) as 
$$
\begin{array}{rcl}
R(f_{1},f_{2}; t)_{-} & \propto & |A_{f_{1}}|^{2}|A_{f_{2}}|^{2}
\exp(-\Gamma |t|) ~ \times \\
&  & \displaystyle\left [ \left (|\xi_{-}|^2 + |\zeta_{-}|^2 \right )
\cosh(y^{~}_D \Gamma t) - 2 {\rm Re} \left (\xi^*_{-}\zeta_{-}\right ) \sinh(y^{~}_D \Gamma t) 
\right . \\
&  & \left .  - \displaystyle\left (|\xi_{-}|^{2}-|\zeta_{-}|^{2}\right ) \cos (x^{~}_D \Gamma t)
+ 2{\rm Im}\left (\xi_{-}^{*}\zeta_{-}\right ) \sin (x^{~}_D \Gamma t) \right ] \; ,
\end{array}
\eqno{\rm (A1)}
$$
and
$$
\begin{array}{rcl}
R(f_{1},f_{2}; t)_{+} & \propto & |A_{f_{1}}|^{2}|A_{f_{2}}|^{2}
\exp(-\Gamma |t|) ~ \times \\
&  & \displaystyle \left [ \frac{ |\xi_{+}|^2 + |\zeta_{+}|^{2} }{\sqrt{1-y^2_D}}
\cosh(y^{~}_D \Gamma |t| + \phi_y) - \frac{ 2{\rm Re} \left (\xi^*_{+}\zeta_{+}\right ) }
{\sqrt{1-y^2_D}} \sinh(y^{~}_D \Gamma |t| + \phi_y) \right . \\
& & \left . \displaystyle - \frac{ |\xi_{+}|^{2}-|\zeta_{+}|^{2} }{\sqrt{1+x^{2}_D}}
\cos \left (x^{~}_D \Gamma |t| +\phi_{x}\right ) + \frac{ 2{\rm Im}\left (\xi^{*}_{+}\zeta_{+}\right ) }
{\sqrt{1+x^{2}_D}} \sin (x^{~}_D \Gamma |t| +\phi_{x}) \right ]  \; ,
\end{array}
\eqno{\rm (A2)}
$$
where the phase shifts $\phi_x$ and $\phi_y$ are defined by
$\tan\phi_{x}= x^{~}_D$ and $\tanh\phi_y = y^{~}_D$ respectively. One can
check that integrating $R(f_{1},f_{2};t)_{C}$ over $t$, where
$t\in (-\infty,+\infty)$, will lead to the time-independent decay rate
$R(f_{1},f_{2})_{C}$ given in eq. (2.21). Eqs. (A1) and (A2) are two basic
formulas for investigating coherent $D^{0}\bar{D}^{0}$ 
decays at asymmetric $\tau$-charm factories.

\vspace{0.3cm}

Another possibility is to measure the time-integrated decay rates
of $(D^0_{\rm phys}\bar{D}^0_{\rm phys})_{C}$ with a proper time cut, which can sometimes 
increase the sizes of $CP$ asymmetries \cite{Xing1}.
In practice, appropriate time cuts can also suppress background and improve
statistic accuracy of signals. If the decay events in the time region $t\in [+t_{0}, +\infty)$
or $t\in (-\infty, -t_{0}]$ are used, where $t_{0}\geq 0$,
the respective decay rates can be defined by
$$
\begin{array}{rcl}
\hat{R}(f_{1},f_{2}; + t_{0})_{C} & \equiv &
\displaystyle\int^{+\infty}_{+t_{0}} R(f_{1},f_{2}; t)_{C} ~ {\rm d}t \; , \\
\hat{R}(f_{1},f_{2}; - t_{0})_{C} & \equiv &
\displaystyle\int^{-t_{0}}_{-\infty} R(f_{1},f_{2}; t)_{C} ~ {\rm d}t \; .
\end{array}
\eqno{\rm (A3)}
$$
By use of eqs. (A1) and (A2), we obtain
$$
\begin{array}{rcl}
\hat{R}(f_{1},f_{2}; \pm t_{0})_{-} & \propto &
|A_{f_{1}}|^{2}|A_{f_{2}}|^{2} \exp(-\Gamma t_{0}) ~ \times \\
&  & \displaystyle \left [ \frac{ |\xi_{-}|^2 + |\zeta_{-}|^{2} }{2\sqrt{1-y^2_D}}
\cosh(y^{~}_D \Gamma t_0 + \phi_y) \mp
\frac{ {\rm Re} \left (\xi^*_- \zeta_{-}\right ) }{\sqrt{1-y^2_D}} 
\sinh(y^{~}_D \Gamma t_0 + \phi_y) \right . \\
& & \left .  \displaystyle - \frac{ |\xi_{-}|^{2}-|\zeta_{-}|^{2} }{2\sqrt{1+x^{2}_D}}
\cos \left (x^{~}_D \Gamma t_{0} +\phi_{x}\right ) \pm \frac{ {\rm Im}\left (\xi^{*}_{-}\zeta_{-}\right ) }
{\sqrt{1+x^{2}_D}} \sin (x^{~}_D \Gamma t_{0} +\phi_{x}) \right ]  \; ,
\end{array}
\eqno{\rm (A4)}
$$
and
\small
$$
\begin{array}{rcl}
\hat{R}(f_{1},f_{2}; \pm t_{0})_{+} & \propto &
|A_{f_{1}}|^{2}|A_{f_{2}}|^{2} \exp(-\Gamma t_{0}) ~ \times \\
&  & \displaystyle \left [ \frac{ |\xi_{+}|^2 + |\zeta_{+}|^{2} }{2 \left (1-y^{2}_D \right )}
\cosh(y^{~}_D \Gamma t_0 + 2 \phi_y) - 
\frac{ {\rm Re} \left (\xi^*_{+} \zeta_{+} \right ) }{1-y^{2}_D}
\sinh(y^{~}_D \Gamma t_0 + 2 \phi_y) \right . \\
& & \left .  \displaystyle - \frac{ |\xi_{+}|^{2}-|\zeta_{+}|^{2} }{2 \left (1+x^{2}_D \right )}
\cos \left (x^{~}_D \Gamma t_{0} + 2\phi_{x}\right ) + \frac{ {\rm Im}\left (\xi^{*}_{+}\zeta_{+}\right ) }
{1+x^{2}_D} \sin (x^{~}_D \Gamma t_{0} + 2\phi_{x}) \right ]  \; .
\end{array}
\eqno{\rm (A5)}
$$
\normalsize
It is easy to check that
$$
\hat{R}(f_{1},f_{2};+0)_{C} + \hat{R}(f_{1},f_{2};-0)_{C}
\; =\; R(f_{1},f_{2})_{C} \; .
\eqno{\rm (A6)}
$$
One can observe that in $\hat{R}(f_{1},f_{2};\pm t_{0})_{C}$ different
terms are sensitive to the time cut $t_{0}$ in different ways. Thus it is possible to
enhance a $CP$-violating term (and suppress the others) via a suitable cut
$t_{0}$.

\vspace{0.3cm}

\begin{flushleft}
\Large\bf Appendix B
\end{flushleft}

In this appendix we take a brief look at the possible effect of $CPT$ violation
in $D^0-\bar{D}^0$ mixing on the decay rates of semileptonic $D$ decays.
For simplicity, we assume the $\Delta Q = \Delta C$ rule and direct $CPT$
invariance in $D$ decays to hold exactly. We also assume the absence of
final-state electromagnetic interactions and other sources of new physics
that could affect the tree-level $W$-mediated $D$ decays.
Due to the presence of $CPT$ violation,
the mass eigenstates $|D_{\rm L}\rangle$ and $|D_{\rm H}\rangle$ can now be expressed as
$$
\begin{array}{lll}
|D_{\rm L} \rangle & = & \displaystyle \cos \frac{\theta}{2} ~ p |D^0\rangle ~ 
+ ~ \sin \frac{\theta}{2} ~ q |\bar{D}^0\rangle \; , \\ \\
|D_{\rm H}\rangle & = & \displaystyle \sin \frac{\theta}{2} ~ p |D^0\rangle ~ 
- ~ \cos \frac{\theta}{2} ~ q |\bar{D}^0\rangle \; , 
\end{array}
\eqno{\rm (B1)}
$$
where $\theta$ is in general complex.
Note that $CPT$ invariance requires $\cos\theta =0$, while
$CP$ conservation requires both $\cos\theta =0$ and $p=q=1$ \cite{Lee-Wu}. 
Taking $\theta = \pi/2$, i.e., $CPT$ symmetry, one can reproduce eq. (2.1) from
eq. (B1). The proper-time evolution of an initially ($t=0$) pure $D^0$ or $\bar{D}^0$ 
turns out to be
$$
\begin{array}{lll}
|D^0_{\rm phys}(t)\rangle & = & \displaystyle \left [ g_{+}(t) + g_-(t) \cos\theta \right ] |D^0\rangle ~ + ~ 
\frac{q}{p} \left [ g_-(t) \sin\theta \right ] |\bar{D}^0\rangle \; ,  \\ \\
|\bar{D}^0_{\rm phys}(t)\rangle & = & \displaystyle \left [ g_{+}(t) - g_-(t) \cos\theta \right ] |\bar{D}^0\rangle ~ + ~ 
\frac{p}{q} \left [ g_-(t) \sin\theta \right ] |D^0\rangle \; ,
\end{array}
\eqno{\rm (B2)}
$$
where $g^{~}_{\pm}(t)$ have been given in eq. (2.6).

\vspace{0.3cm}

Starting from eq. (B2), one can calculate the $CP$ asymmetry $\bar{\Delta}_D$ defined in
eq. (3.5) for semileptonic $D$ transitions. We find
$$
\bar{\Delta}_D \; = \; \frac{2 x^{~}_D \alpha {\rm Im}(\cos\theta) \; +\; 2 y^{~}_D {\rm Re}(\cos\theta) }
{(1+ \alpha ) \; +\; (1- \alpha ) |\cos\theta|^2} \; ,
\eqno{\rm (B3)}
$$
where $\alpha = (1-y^2_D)/(1+x^2_D)$ has been defined before.
Clearly $\bar{\Delta}_D =0$, if there is no $CPT$ violation in $D^0-\bar{D}^0$ mixing 
(i.e., $\cos\theta =0$). Since $|\cos\theta|$ must be a small quantity, the $|\cos\theta|^2$ term 
in the denominatior of $\bar{\Delta}_D$ is negligible. Anyway observation of the signal
$\bar{\Delta}_D$ will be greatly difficult in practice, since its magnitude is 
suppressed by both the small mixing rate and the small $CPT$ asymmetry.

\vspace{0.3cm}

Next let us assume the experimental scenario to be an asymmetric $\tau$-charm factory,
in which $D^0\bar{D}^0$ pairs can be coherently produced at the $\psi (3.77)$ or
$\psi (4.14)$ resonance and the time-dependent measurements of their decays are available.
To probe possible $CPT$ violation in $D^0-\bar{D}^0$ mixing, we 
consider the case that one $D$ meson decays to the semileptonic state $e^{\pm}X^{\mp}_e$ 
at (proper) time $t_e$ and the other to the semileptonic state $\mu^{\mp}X^{\pm}_{\mu}$ 
at $t_{\mu}$. The joint decay rate for having such an event can be given as a function of the
decay-time difference $t\equiv t_{\mu} - t_e$. For simplicity and definition,
we choose $t>0$ by convention. This implies that $e^{\pm}X^{\mp}_e$ events 
may serve for flavor-tagging of $\mu^{\mp} X^{\pm}_{\mu}$ events. After a lengthy calculation, 
we obtain 
$$
\begin{array}{lll}
R(e^{\pm}X^{\mp}_e, \mu^{\mp}X^{\pm}_{\mu}; t)_- & \propto & |A_e|^2 |A_{\mu}|^2
\exp (-\Gamma t) \left [ \cosh (y^{~}_D \Gamma t) ~ + ~ \cos (x^{~}_D \Gamma t) \right . \\
&  & \left . \pm ~ 2 {\rm Re}(\cos\theta) \sinh (y^{~}_D \Gamma t) ~ \pm ~ 2 {\rm Im}(\cos\theta)
\sin (x^{~}_D \Gamma t) \right ] \; , 
\end{array}
\eqno{\rm (B4)}
$$
and
$$
\begin{array}{ll}
& R(e^{\pm}X^{\mp}_e, \mu^{\mp}X^{\pm}_{\mu}; t)_+ \; \propto \; \displaystyle |A_e|^2|A_{\mu}|^2
\exp (-\Gamma t) \left \{ \frac{\cosh(y^{~}_D \Gamma t + \phi_y)}{\sqrt{1-y^2_D}} + 
\frac{\cos(x^{~}_D \Gamma t + \phi_x)}{\sqrt{1+x^2_D}} \right . \\
& ~~~~~~~~~~~ \displaystyle \pm \frac{2|\cos\theta|}{\sqrt{x^2_D +(2-y^{~}_D)^2}} 
\left [ \cos(\Theta +\omega_{-}) \exp(+y^{~}_D \Gamma t)
 -  \cos(\Theta +\omega_{+} +x^{~}_D \Gamma t) \right ] \\
& ~~~~~~~~~~~ \displaystyle \left . \mp \frac{2|\cos\theta|}{\sqrt{x^2_D +(2+y^{~}_D )^2}} 
\left [ \cos(\Theta -\omega_{+}) \exp(-y^{~}_D \Gamma t)
 - \cos(\Theta -\omega_{+} -x^{~}_D \Gamma t) \right ] \right \} \; ,
\end{array}
\eqno{\rm (B5)}
$$
where $\phi_x$ and $\phi_y$ have been defined in Appendix A, and the phase shifts 
$\omega_{\pm}$ and $\Theta$ are
defined by $\tan\omega_{\pm}\equiv x^{~}_D /(2\pm y^{~}_D )$ and
$\tan \Theta \equiv {\rm Im}(\cos\theta)/{\rm Re}(\cos\theta)$ respectively. 
In obtaining eqs. (B4) and (B5), we have neglected those higher-order terms of
$\cos\theta$. It is clear that the opposite-sign dilepton events 
$R(l^{\pm}X^{\mp}_l, l^{\mp}X^{\pm}_l; t)_{C}$ cannot be used to explore possible $CPT$ violation in
$D^0-\bar{D}^0$ mixing, because the time order of $l^+$ and $l^-$ is hardly distinguishable 
in practical experiments. In addition, the signal of $CPT$ violation cannot manifest itself 
in the time-integrated decay rates of $(D^0_{\rm phys}\bar{D}^0_{\rm phys})_C \rightarrow
(e^{\pm}X^{\mp})(\mu^{\mp}X^{\pm})$, as obviously shown by the equations above. That is
why we need an asymmetric $\tau$-charm factory to test $CPT$ symmetry in $D^0-\bar{D}^0$ mixing.

\vspace{0.3cm}

Of course, $CPT$ violation can appear in many other decay modes of neutral $D$ mesons.
The semileptonic processes discussed above are more attractive to us for the study of 
$CPT$ violation, since they do not involve $CP$ asymmetry
in $D^0-\bar{D}^0$ mixing (measured by $|q/p|\neq 1$) and other $CP$-violating signals.
In general, however, both direct and indirect $CPT$ asymmetries as well as 
$\Delta Q = - \Delta C$ transitions (and other sources of new physics) are possible to
affect the decay modes in question \cite{Xingnote}.

\end{document}